\documentclass[aps,preprint,showpacs,preprintnumbers,amsmath,amssymb,longbibliography]{revtex4-1}
\usepackage{graphicx}
\usepackage{epstopdf}
\usepackage{dcolumn}
\usepackage{bm}
\usepackage{color}
\def\degree{\ensuremath{^{\circ}}}

\begin{document}

\preprint{APS/123-QED}

\title{The multi-scale nature of Wall shear stress fluctuations in turbulent Rayleigh-B\'enard convection}

\author{Christoph Bruecker}
 \email{christoph.bruecker@city.ac.uk}
 \affiliation{
School of Mathematics, Computer Science and Engineering, City University London, UK}

\author{R.~du~Puits}
 \affiliation{
Institute of Thermodynamics and Fluid Mechanics, Technische Universitaet Ilmenau, POB 100 565, 98684 Ilmenau, Germany
}%

\date{\today}

\begin{abstract}
Measurements of wall shear-stress fluctuations on very long timescales ($\ge$ 1900 free-fall time units) are reported for turbulent Rayleigh-Benard (RB) convection in air at the heated bottom plate of a RB cell, 2.5 m in diameter and 2.5 m in height. The novel sensor simultaneously captures the fluctuations of the magnitude and the direction of the wall shear stress vector $\boldsymbol{\tau}(t)$ with high resolution in the slow air currents. The results show the persistence of a tumble-type structure, which is in a bi-stable state as it oscillates regularly about a mean orientation at a timescale that compares with the typical eddy turnover time.  The mean orientation can persist almost hundreds of eddy turnovers, until a re-orientation of this structure in form of a slow precession sets in, while a critical weakening of the mean wall shear stress magnitude - respectively the mean wind - is observed. The amplitudes of turbulent fluctuations in the streamwise wall shear-stress $\tau_x$ along mean wind direction reveal a highly skewed Weibull distribution, while the fluctuations happening on larger time scales follow a symmetric Gaussian distribution. Extreme events such as local flow reversals with negative $\tau_x$ are recovered as rare events and correlate with a rapid angular twist of the wall shear-stress vector. Those events - linked to critical points in the skin friction field - correlate with  the coincidence of signals at the tails in both probability distributions.

\end{abstract}

\maketitle

\section{\label{sec:Introduction}Introduction}

The heat transport by turbulent convection is characterized by a subtle interplay between small-scale turbulence in the near-wall flow field at the heated/cooled surfaces and the large-scale structures evolving or being present far from the wall. In spite of considerable experimental and theoretical effort, the spatial and temporal dynamics of these large-scale structures and their dependency of the width-to-height ratio of the adjacent fluid layer are not very well understood. Since these large-scale structures also affect the flow field adjacent to the wall, they have to be considered too, studying the convective heat transport from a solid surface to a surrounding fluid.
Ludwig Prandtl was the first person, who linked heat and momentum transport throughout a convective boundary layer in his so-called mixing length theory \cite{Prandtl1925}. The core idea of this theory is that turbulent parcels of fluid transport momentum and heat simultaneously. Therefore, Prandtl believed that both transport coefficients are equal. Today, we know that this is not fully true. But, following Prandtl's idea, it is undoubtedly that the local flow field close to the wall determines the heat transport from/toward a heated/cooled surface. In the work presented here, we focus on measurements of the local wall shear stress, which has been identified as a quantity that is directly linked to the local heat transport (if the fluid layer adjacent to the wall is not laminar and convection dominates over diffusion).

The link between the local heat transport and the wall shear stress in a non-laminar boundary layer has been quantitatively studied first by Ludwieg \cite{Ludwieg1956}. He experimentally determined the ratio between the transport coefficients for heat and momentum in a fully developed turbulent pipe flow and, unlike Prandtl, he found this ratio to be different from one. Recent experiments using micro-pillar wall shear stress sensors in a turbulent boundary layer flow along a flat plate showed the existence of singularities in the wall shear stress vector field. For the first time, those measurements highlighted the importance of the directional information of the wall shear stress vector and its topology \cite{Bruecker2015}. It is the authors conclusion that both wall shear stress components are essential to draw correct conclusions about the correlation between $\tau_W$ and the local heat transport from the wall. In general, time-resolved measurements of the wall shear stress in thermal convection are rare, and in particular, measurements in turbulent Rayleigh-B\'enard (RB) convection, a fluid layer heated from below and cooled from the top, are completely missing. The current status quo in such data knowledge is only provided by recent highly resolved Direct Numerical Simulations (DNS). Those simulations can provide the local wall shear stress vector information in time, but only for a limited simulation periods. First wall shear stress data for a similar RB convection flow as studied herein shows also the existence of singularities in the wall shear stress vector field similar as those reported in \cite{Bruecker2015}, which were found to be footprints of large eruptions of fluid parcels from the wall \cite{Bandaru2015}. However, the state-of-the-art computational power allows only to run the simulations for short periods in time and are limited to Rayleigh numbers as low as $\textrm{Ra}=10^{10}$~(see e.g. \cite{Scheel2014}).

Recently (in 2017), Bruecker and Mikulich developed a novel sensor particularly designed to measure the wall shear stress in low-speed air flows \cite{Bruecker2017}. As reported in \cite{Bruecker2017}, its sensitivity and dynamic response were sufficiently good to apply it in the so-called Barrel of Ilmenau (BOI), a large-scale RB experiment, which will be described in detail below. The flow in such a convection cell is known to generate a large-scale circulation (LSC), whose shape mainly depends on the diameter-to-height ratio of the test cell. This ratio defined as $ \Gamma=D/H$, is commonly referred to as aspect ratio. The LSC exhibits the shape of a single roll for aspect ratios of order unity, while for larger or smaller aspect ratios a pattern of multiple rolls evolves \cite{dupuits07b,Reeuwijk08a,Bailon10,Mishra11}. It is already known that the single roll structure exhibits a number of various flow modes, like e.g. the oscillation of its plane around the mean \cite{Resagk06}, the torsional mode disentangling the upper and the lower part of the flow structure \cite{Ahlers09}, the sloshing mode \cite{Xi09}, and various kinds of re-orientations like reversals \cite{Sreenivasan2002}, rotations or cessations \cite{Brown2006}. Independent on its particular mode, the LSC forms distinct boundary (shear) layers near the top and bottom plates as well as near the sidewalls. The wall shear stress at the surfaces of the plates and the sidewall reflects both the particular shape and the dynamics of the LSC.

The present work reports the first application of this sensor in the BOI, which addresses the hitherto unknown long-term dynamics of the wall shear stress field by simultaneously measuring the magnitude and direction of the wall shear stress vector. The results display the long-term behaviour of the local wall shear stress in turbulent RB convection and give insight into the long-term statistics and dynamics of the flow pattern, permitting validation of recent DNS. The sensor has the appropriate temporal resolution along with a very high sensitivity to capture singularities in the wall shear stress field at long sampling periods of the order of hours and more. Furthermore, the system also allows to recover the dynamics of the LSC \cite{Hansen1992,Sreenivasan2002,Funfschilling2004,Araujo2005, Brown2006,duPuits2007}. Hence, our wall shear stress measurements contribute to a better understanding, how the emission of strong plumes from the boundary layer is coupled with the global flow field \cite{Villermaux1995,Qiu2000,Kadanoff2001,Xi2004}.

\section{\label{sec:Setup}Experimental set-up and measurement technique}

\subsection{\label{sec:BOI}The large-scale Rayleigh-B\'enard Experiment ``Barrel of Ilmenau''}

\begin{figure*}
\includegraphics[width=14cm]{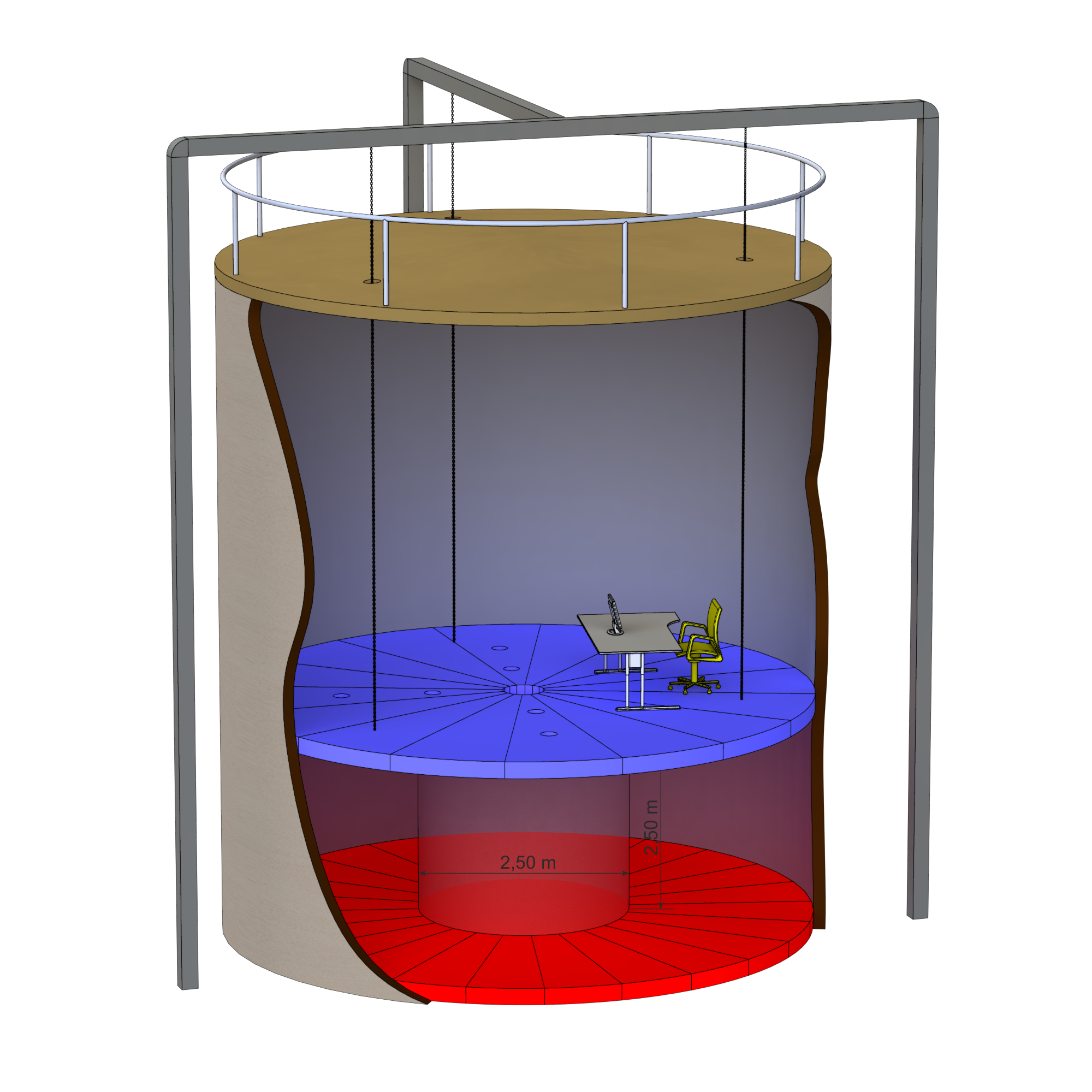} 
\caption{\label{fig:BOI}Sketch of the large-scale Rayleigh-B\'enard experiment ``Barrel of Ilmenau'' with the smaller inset of $D=2.5~\textrm{m}$. The origin of a Cartesian coordinate system is fixed with the centre of the bottom plate (the location of the wall shear stress sensor) in the $x,y$ plane and the $z$ axis pointing normal to the wall towards the top plate.}
\end{figure*}

The wall shear stress measurements were carried out in the so-called ``Barrel of Ilmenau (BOI)'' (see Fig.~\ref{fig:BOI}) with the sensor mounted at the center of the bottom plate. The BOI represents a classical Rayleigh-B\'enard (RB) experiment using air ($Pr=0.7$) as working fluid. It is confined in a well-insulated container of cylindrical shape with an inner diameter of $D=7.15$~m. A heating plate at the lower side releases the heat to the air layer, and a cooling plate at the upper side removes it. Both plates are carefully levelled perpendicular to the vector of gravity with an uncertainty of less than 0.15~degrees. The thickness of the air layer $H$ can be varied continuously between $6.30~\textrm{m}>H>0.15~\textrm{m}$ by moving the cooling plate up and down. The temperature of both plates can be set to values between $20~\degree\textrm{C}<T_h<80~\degree\textrm {C}$ (heating plate) and $10~\degree\textrm{C}<T_c<30~\degree\textrm{C}$ (cooling plate). Due to the specific design of both plates (for details see \cite{dupuits2013}), the temperature at their surfaces is very uniform and the deviation does not exceed 1.5~\% of the total temperature drop $\Delta T=T_h-T_c$ across the air layer. The variation of the surface temperature over the time is even smaller and remains below $\pm 0.02~\textrm{K}$. The sidewall of the model room is shielded by an active compensation heating system to inhibit any heat exchange with the environment. Electrical heating elements are arranged between an inner and an outer insulation of 16~cm and 12~cm thickness, respectively. The temperature of the elements is controlled to be equal to the temperature at the inner surface of the wall, and thus, a heat flux throughout the side-wall is impossible. As described in Section~\ref{sec:Measurement} the top-plate has several glass windows, which allow optical access to the inner of the test section. Our measurements were undertaken inside a smaller inset of $D=2.5~\textrm{m}$ and $H=2.5~\textrm{m}$ (see Fig.~\ref{fig:BOI}). This reduces the Rayleigh number, $\textrm{Ra}=(\beta g\Delta TH^3)/(\nu\kappa)$ with being $\beta$ the thermal expansion coefficient, $g$ the gravitational acceleration, $\nu$ the kinematic viscosity, and $\kappa$ the thermal diffusivity, to $Ra=1.58\times10^{10}$, and allows to compare our results with currently running DNS.

\begin{figure}
	\includegraphics[width=10cm]{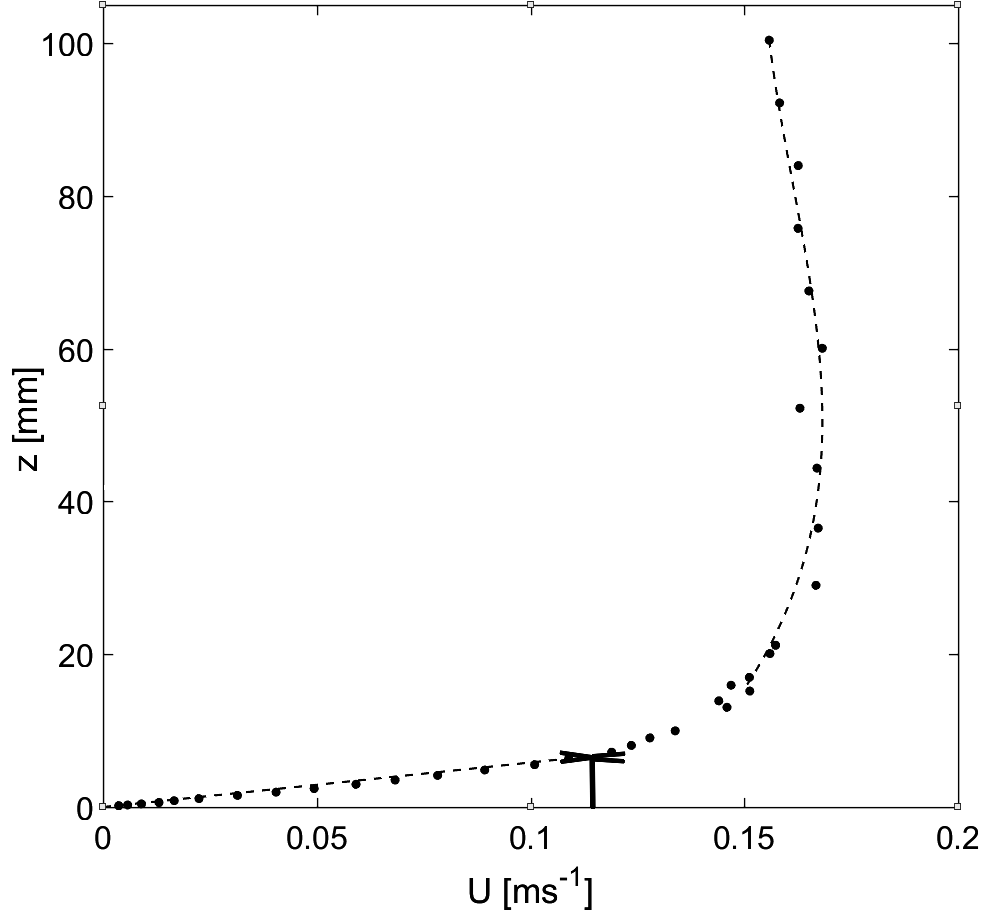} 
	\caption{Mean velocity profile in the boundary layer of the barrel of Ilmenau. Inserted is a true-scale sketch of the sensor with its head at $z_0=7~\textrm{mm}$, illustrating that it is fully surrounded by the linear part of the velocity profile.}
	\label{fig:profile}
\end{figure}

The characteristic timescale of the flow is the so-called free-fall time unit, defined as $T_f=\sqrt{H/g\beta\Delta T}$, which is about $T_f=2.73~\textrm{s}$ for the current configuration at a temperature difference of $\Delta T = 10~\degree\textrm{C}$. Another timescale is the characteristic eddy turnover time of the LSC, which is calculated from the mean wind $U=0.15~\textrm{m/s}$ and the circumference of the cell to about $T_e=50~\textrm{}s$.
Previous flow studies in the BOI using Laser Doppler Velocimetry show a typical boundary layer thickness of the order of $\delta =50~\textrm{mm}$. The smallest turbulent scales are about 30~mm and correlate to the highest reported frequencies of the order of 2~Hz (\cite{Resagk06, dupuits07a}). A typical profile of the mean velocity at the position of the sensor, measured using Laser Doppler Velocimetry, is shown in Fig.~\ref{fig:profile}. The linear part of the profile as indicated by the dashed line represents the viscous sublayer close to the wall. Measurements by Ampofo and Karayiannis \cite{Ampofo03} in a similar low-turbulence convection flow as studied herein show that the viscous sublayer thickness is of order of 10~\% of the outer boundary layer, similar as observed in the BOI. For subsequent discussion, the picture additionally displays a true-scale sketch of the sensor. This illustrates, that the sensor is one order of magnitude smaller than the typical boundary layer thickness in the BOI.

\subsection{\label{sec:Sensor}The wall shear stress sensor}

The sensor including its calibration in the BOI is described in detail in \cite{Bruecker2017} . The interested reader is referred to the reference, while we give here only a short review. The sensor resembles an artificial wind receptor hair with a stem, which is bonded at the wall with the foot in a flexible membrane. It tilts with the flow like an inverted pendulum. The sensor's head consists of a pappus of slender hairs with a diameter of a few tens of microns. They act as an antenna maximizing the drag force of an imposing flow. The mechanical behavior of the sensor is described in Bruecker and Mikulich as a second-order harmonic oscillator in overdamped condition \cite{Bruecker2017}. A calibration of the mechanical model can provide the two unknown variables of the solution to the response function, the constant gain $K$ and the cut-off frequency $f_c$, the frequency at which the sensor can no longer follow the excitation (the response starts to roll-off with -20 dB per decade).

\begin{figure}
	{\includegraphics[width=14cm]{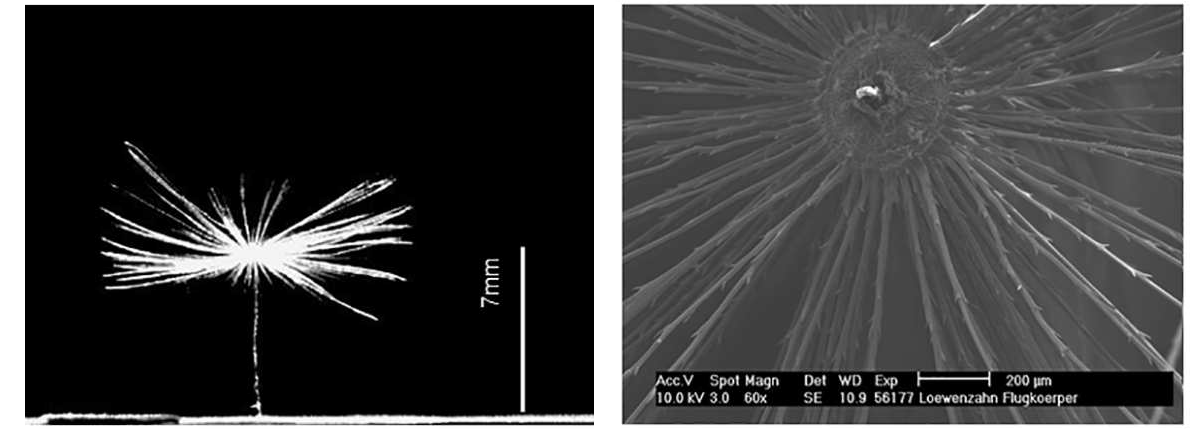}}
	\caption{Pictures of the pappus sensor fixated with the stem at the bottom plate of the BOI. Pictures show the side view (left) with a scale bar and the top view with higher magnification (right). Taken from \cite{Bruecker2017}.}
	\label{fig:Pappus}
\end{figure}

\begin{figure}
	{\includegraphics[width=14cm]{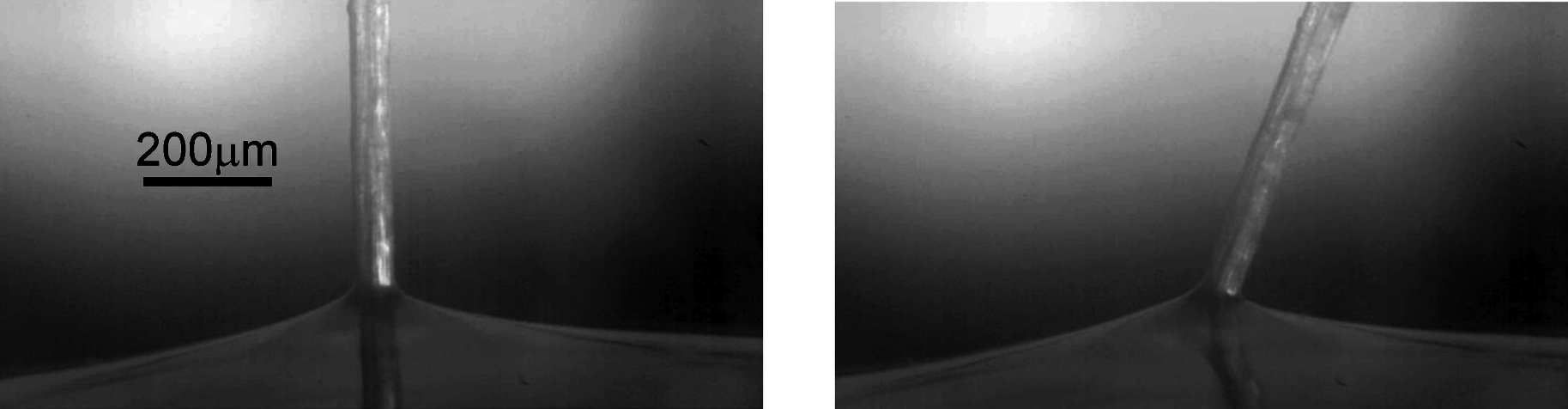}}
	\caption{Close-up view of the flexible foot fixation of the stem at the bottom plate of the BOI in a thin Silicone rubber membrane. Sensor at rest (left) and under strong wind load (right). Note that the stem keeps its straight shape under load. Taken from \cite{Bruecker2017}.}
	\label{fig:Pappus_in_Silicone}
\end{figure}

A detailed view of the sensor is shown in Fig.~\ref{fig:Pappus}. The stem and head were taken from a nature-grown Dandelion \cite{Bruecker2017} with a pappus consisting of a total 86 radially arranged slender hairs (mean length $l=7~\textrm{mm}$, mean diameter $d=30~\mu\textrm{m}$. It has a stem height of $z_0=7~\textrm{mm}$ and the overall radial diameter of the pappus is about $D_p=14~\textrm{mm}$. The Reynolds number $\textrm{Re}$ of the flow around the individual hairs - simplified as thin cylinders of diameter $d$ - is of the order of $\textrm{Re}_d\approx2$ for air speeds of $1~\textrm{m/s}$. Thus, the drag is dominated by viscous friction and scales, therefore, approximately linear with flow speed  \cite{Liebe2006,Pandolfi2013,Casseau2015}. The elastic membrane, at which the stem's foot is bonded, is made from rubber silicone (Polydimethylsiloxane, PDMS; Youngs modulus $E\approx1.5~\textrm{MPa}$) and acts as a torsional spring with uniform bending stiffness in radial direction. A close-up view of the membrane with the stem is given in Fig.~\ref{fig:Pappus_in_Silicone}. When the stem with the pappus is exposed to an air flow parallel to the wall, the resulting torque tilts the stem around the membrane. As the tilt is proportional to the torque, the latter can be measured indirectly by the end-to-end shift vector $\textbf{Q}(t)$ of the tip relative to the wind-off reference. For small tilt angles this is approximately the projection of the vector in the horizontal x-y plane at $z=z_0$ with $\textbf{Q}(t) =   \left( Q_x(t), Q_z(t) \right)  $, which we capture by imaging the sensor head's orbital motion from top. 

At a sufficiently small sensor scale with $z_0\ll \delta$, $\textbf{Q}(t)$ is directly proportional to the wall shear stress vector $\boldsymbol{\tau}(t)= \left(  \tau_x(t), \tau_y(t) \right) $, see also \cite{Bruecker2014}. Assuming the mean flow is parallel to the wall in x-direction the streamwise component is defined by the wall-normal velocity gradient $\partial u_x/\partial z|_{z=0\textrm{mm}}$ and the spanwise component is $\partial u_y/\partial z|_{z=0\textrm{mm}}$ respectively. Using a Taylor expansion, the information of the velocity field in the $x-y$ plane close to the wall at a distance $z=z_0$ is related to these wall shear-stress components as follows: 

\begin{equation}\label{eq:wall_stress2}
\tau_{x,y} = \mu~\frac{u_{x,y}(z_0)}{z_0}+\mathrm{O} (z_0)^2
\end{equation} 

For small wall distances $z_0\ll \delta$, the second order term in Eq.~(\ref{eq:wall_stress2}) can be neglected.  This approximation is valid close to the wall within the viscous sublayer of a turbulent boundary layer (TBL), see the discussion above. The low Re number flow around the micron-size pappus hairs at the sensor head suggests a linear relationship between the pappus drag and the air velocity \cite{Liebe2006,Pandolfi2013,Casseau2015}, therefore we can approximate the sensor tilt proportional to the velocity as follows: 

\begin{equation}\label{eq:wall_stress3}
u_{x,y}(z_0) \approx ~K \cdot Q_{x,y}
\end{equation}

Therefore, once the tip displacement of the sensor is calibrated with respect to a reference flow, the  proportionality constant $K$ in Eq.~(\ref{eq:wall_stress3}) can be determined and the gain of the mechanical system is given. Using Eq.~(\ref{eq:wall_stress2},\ref{eq:wall_stress3}) for a given fluid viscosity $\mu$ then allows to calculate the time-dependent vector $\boldsymbol{\tau}(t)$ in magnitude and direction from time-resolved measurements of $\textbf{Q}(t)$ during the motion of the sensor.

\subsection{\label{sec:Measurement}Optical set-up for sensor imaging}
\begin{figure}
	\includegraphics[width=14cm]{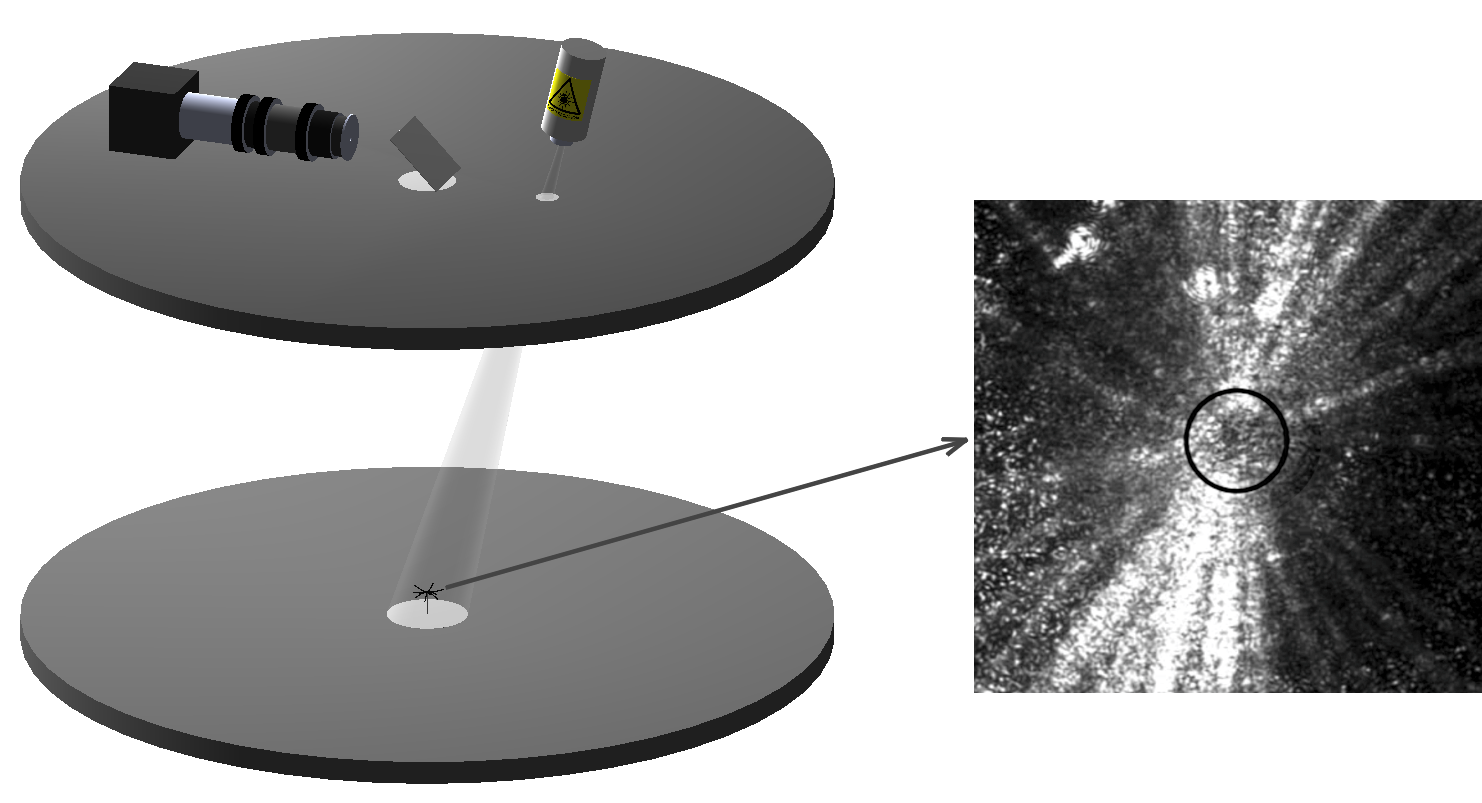} 
	\caption{Optical set-up of the wall shear stress measurement technique applied at the hot bottom plate (left). Top view of the pappus sensor, the circle marks the center of the sensor head (right).}
	\label{fig:setup}
\end{figure}

Fig.~\ref{fig:setup} shows the optical set-up of our measurements. The pappus sensor at the bottom plate was illuminated by a defocused Laser beam (Raypower 5000, 5~W power at $\lambda=532~\textrm{nm}$, Dantec Dynamics, Skovlunde, Denmark) through a glass window in the top plate . The diameter of the laser beam is about 1.5~mm at the aperture of the laser and it is further expanded to illuminate a spot of 50~mm diameter at the floor. A CCD camera (mvBlueFOX3-1031, Matrix Vision, Oppenweiler, Germany) placed on top of the cooling plate acquires the deflection of the sensor head in the wall-parallel $x,y$ plane with a resolution of 2048~x~1536~px$^2$ and a frame rate of 10~Hz. The camera is equipped with a long-distance microscope (model K2/SC\texttrademark,  Infinity Photo-Optical, Goettingen, Germany), which provides a resolution of 185~px/mm. A total number of 54,000 images was recorded in a single measurement campaign. The images are streamed via USB~3 to the hard disc of a desktop. This equates to a maximum of 1.5~hours of observation time per experiment. To avoid any vibrations during the recordings, the facility was left alone after starting the recording and no external disturbance could enter the RB cell. In order to remove any vibration induced by leaving and re-entering the facility, the first and the last 2-3~minutes were rejected when analysing the data.

The tip displacement vector in the images is obtained using a 2D cross-correlation method similar as in Particle Image Velocimetry technique \cite{Raffel2007}, where we compare a quadratic subsection of the sensor image between wind-off and wind-on situation. The shift in tip position relative to wind-off is determined with subpixel accuracy using a 2D Gaussian fit of the correlation peak, which has an uncertainty of about 0.05 pixel. A reference marker on the floor is used to correct for potential vibrations of the camera during the recordings. After multiplication of the shift with the lens magnification, the vector $\textbf{Q}(t)$ of the sensor head is recovered for each time-step in the image sequence. With the obtained gain $K$ from calibration and $z_0$ as the known sensor wall-normal distance, $\boldsymbol{\tau}(t)$ is determined according Eq.~(\ref{eq:wall_stress2},\ref{eq:wall_stress3}).

\subsection{\label{sec:Calibration}In-situ calibration}

The sensor represents a mechanical system that can be modeled as a second-order harmonic oscillator in an overdamped condition, see Bruecker and Mikulich \cite{Bruecker2017}. A calibration provides the constant gain $K$ and the cut-off frequency $f_c$, at which the sensor can no longer follow the signal (the response starts to roll-off at -20 dB per decade). The latter quantity is inverse to the response time, which can be measured in a step-response test. Therefore, the stem was tilted away from its resting position and the head was recorded while flexing back after unloading. The response time was measured to about  $\tau_{95}=10~\textrm{ms}$. Hence, the sensor can follow the fluctuations up to frequencies of 100~Hz with a constant amplitude response.

\begin{figure}
 \includegraphics[width=12cm]{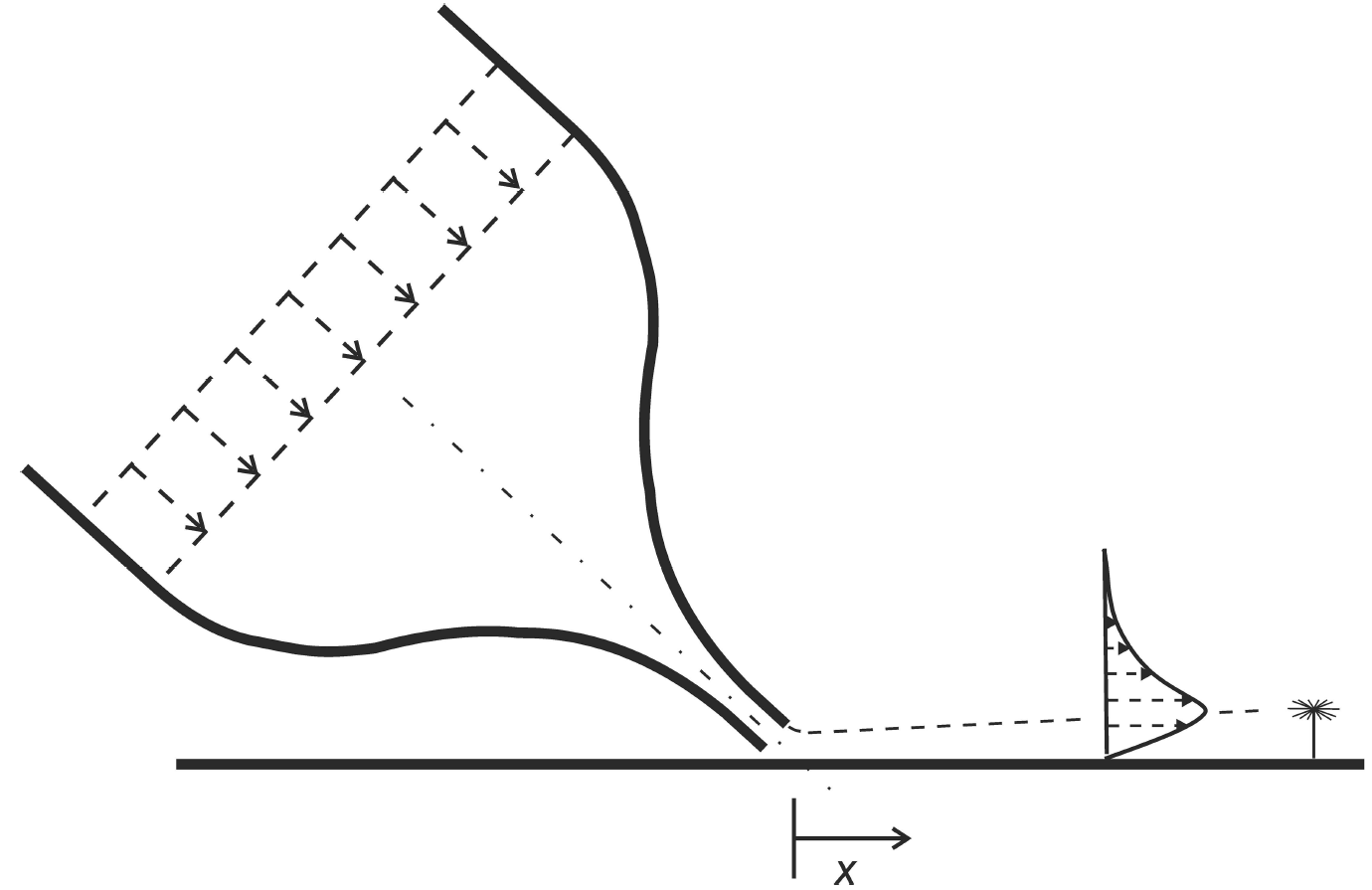}
 \caption{Sketch of the wall-jet apparatus. Air flow is generated with a planar nozzle flow that is tilted at an angle of $45~\textrm{degree}$ towards the bottom heated plate of the research facility and generates a Blasius-type wall-jet in direction of the sensor. The sensor is located at the centre of the plate and the nozzle exit is at a radial offset of 20 slot heights $h_s$ away. The flow profile at the sensor location is well-defined as the laminar Blasius wall-jet.}
\label{fig:Wall_Jet_Apparatus}
\end{figure}

The gain was measured in-situ using a wind-generating device placed inside the BOI under isothermal conditions. The reader is again referred to the details given in Bruecker and Mikulich \cite{Bruecker2017}. The device resembles a small Eiffel-type wind tunnel \cite{Levin2005} with an open rectangular nozzle (width $w_{s}=120~\textrm{mm}$, height $h_{s}=5~\textrm{mm}$). It is arranged in front of the sensor at the centre of the heating plate at a distance of $20 h_{s}$ (see Fig.~\ref{fig:Wall_Jet_Apparatus}). A centrifugal fan generates the flow in the device and guides the air through the smooth contraction unit (ratio 20:1) to the exit slot. The outcoming flow generates a well-defined laminar Blasius-type wall-jet \cite{Glauert1956,Gogineni1997,Levin2005,Bajura1970,Lai1996} along the bottom wall of the barrel. Since the aspect ratio of the slot is high, the flow in the centre plane remains nearly two-dimensional until far distances from the exit.

 Five different jet velocities $v=0.16~\textrm{m/s}$, $v=0.40~\textrm{m/s}$, $v=0.75~\textrm{m/s}$, $v=1.00~\textrm{m/s}$ and $v=1.50~\textrm{m/s}$ have been set and the shift vector $\textbf{Q}$ of the sensor head was measured. These were repeated with the sensor facing the wind from four orthogonal directions. The results of the calibration demonstrate that the direction of $\textbf{Q}$ is always aligned with the jet flow axis and the magnitude $||\textbf{Q}||$ increases approximately linear with the jet velocity at $z_0$ up to $v=1.0~\textrm{m/s}$. Beyond this, the recordings at  $v=1.5~\textrm{m/s}$ show that the configuration of the pappus' hairs starts to change over time and the linear relationship is no longer valid. In the convection flow studies in the barrel the expected maximum velocities are about $1~m/s$ and thus, a linear regression formula can be applied for the calibration in the velocity range of $0 - 1~\textrm{m/s}$, recalling that a linear relationship is expected between air velocity and pappus drag \cite{Liebe2006,Pandolfi2013,Casseau2015} as observed in nature for typical wind speeds of $\le 1~\textrm{m/s}$. The linear regression of the data results to a value of $K=1000~\textrm{s}^{-1}$. As an example, a measured head shift of $500~\mu\textrm{m}$ represents a velocity of $u(z_0)=0.5~\textrm{m/s}$.  
\subsection{\label{sec:Error}Sensor resolution and data processing}
The so calibrated sensor at the centre of the bottom plate is capable to resolve the near-wall velocity $u_ {x,y}(z=z_0)$  from minimum velocities of the order of $0.4 \times 10^{-3}~\textrm{m/s}$ (defined by the uncertainty in optical detection of the shift vector $\textbf{Q}$) up to maximum velocities of $1~\textrm{m/s}$ (limited by the linear regression curve). This is the typical range of air velocities in the barrel that appear in the vicinity of the heated bottom plate. The standard error in the linear regression is about $4 \times 10^{-3}~\textrm{m/s}$. The typical timescales of the velocity flucutations in the BOI, measured in earlier experiments using LDA \cite{duPuits2007} are of the order of 0.5~s (2Hz), which is also well below the sensor's critical cut-off frequency of 100Hz. The present data have been acquired with a sampling frequency of 10Hz. I order to remove spurious samples, the data was filtered in time with a 4th order Butterworth low-pass filter designed with a -3~dB cutoff frequency. We consider this as the optimal configuration to capture the full dynamics of the wall shear stress fluctuations in the convective airflow in the BOI. The spatial resolution of the sensor is defined by the radius of the pappus which amounts to about 7mm. The LDA data mentioned above \cite{duPuits2007} show that the typical small-scale structures have a minimum size of the order of $30~\textrm{mm}$. Therefore, along with a high sensitivity, the sensor also provides a sufficiently well spatial resolution to map the full dynamics of the flow also in the smallest scales. The arguments given above show, that the described sensor was particularly specially designed for the conditions in the BOI which is a very slow, but turbulent air flow. Presently, there is no other sensor capable to measure the wall shear stress and it's fluctuations under these conditions, which was the motivation for the sensor development in 2017 \cite{Bruecker2017}.

In order to make our data comparable with results obtained in recent PIV measurements, we consider in the following the components of the viscosity-divided wall shear stress $\tau_{x,y} / \mu$ with
  \begin{equation}\label{eq:WS}
  \tau_x(t) / \mu = C Q_x(t) / z_0, ~~~~   \tau_y(t) /  \mu = C Q_y(t) / z_0
  \end{equation}
  and we define the direction and the magnitude as follows:
  \begin{equation}\label{eq:angle_magnitude}
  \Phi(t) = \textrm{arctan}~\frac{\tau_y(t)}{\tau_x(t)} \hspace{2cm} \Psi(t)= \frac{1}{\mu}||\tau|| = \frac{1}{\mu} \sqrt{\tau_x^2(t)+\tau_y^2(t) }
  \end{equation}

\section{\label{sec:Results}Results and Discussion}

The wall shear-stress fluctuations at the surface of the bottom plate reflect the dynamics of the convective boundary layer exhibiting eruptions of plumes overlaid with the dynamics of the mean wind  driven by the LSC. In order to recover the large-scale motion from the signals, we applied another low-pass filter on the recorded data (4th order Butterworth low-pass filter designed with a -3~dB cutoff frequency at 0.003~Hz). Based on the notation used in Shi et al. \cite{Shi12}, we define the large-scale direction and magnitude of the wall shear stress based on the low-pass filter as $\widetilde{\Phi}_{LSC}$ and  $\widetilde{\Psi}_{LSC}$. Their values indicate the direction and the magnitude of the mean wind that is imposed by the LSC. This is useful to analyse the wall shear stress data in a plane, which is aligned with the instantaneous direction of the LSC in the cell.

\begin{figure}
\includegraphics[width=16cm]{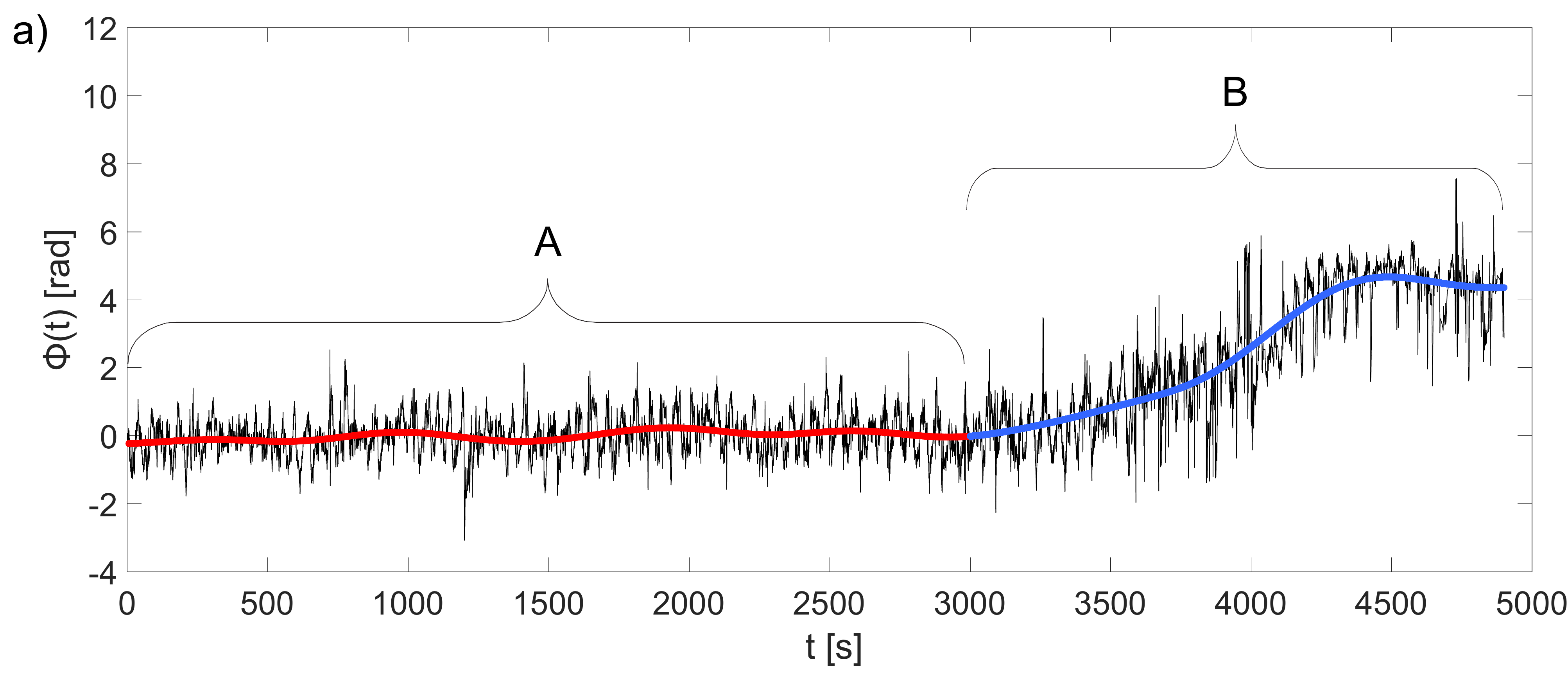} 
\includegraphics[width=16cm]{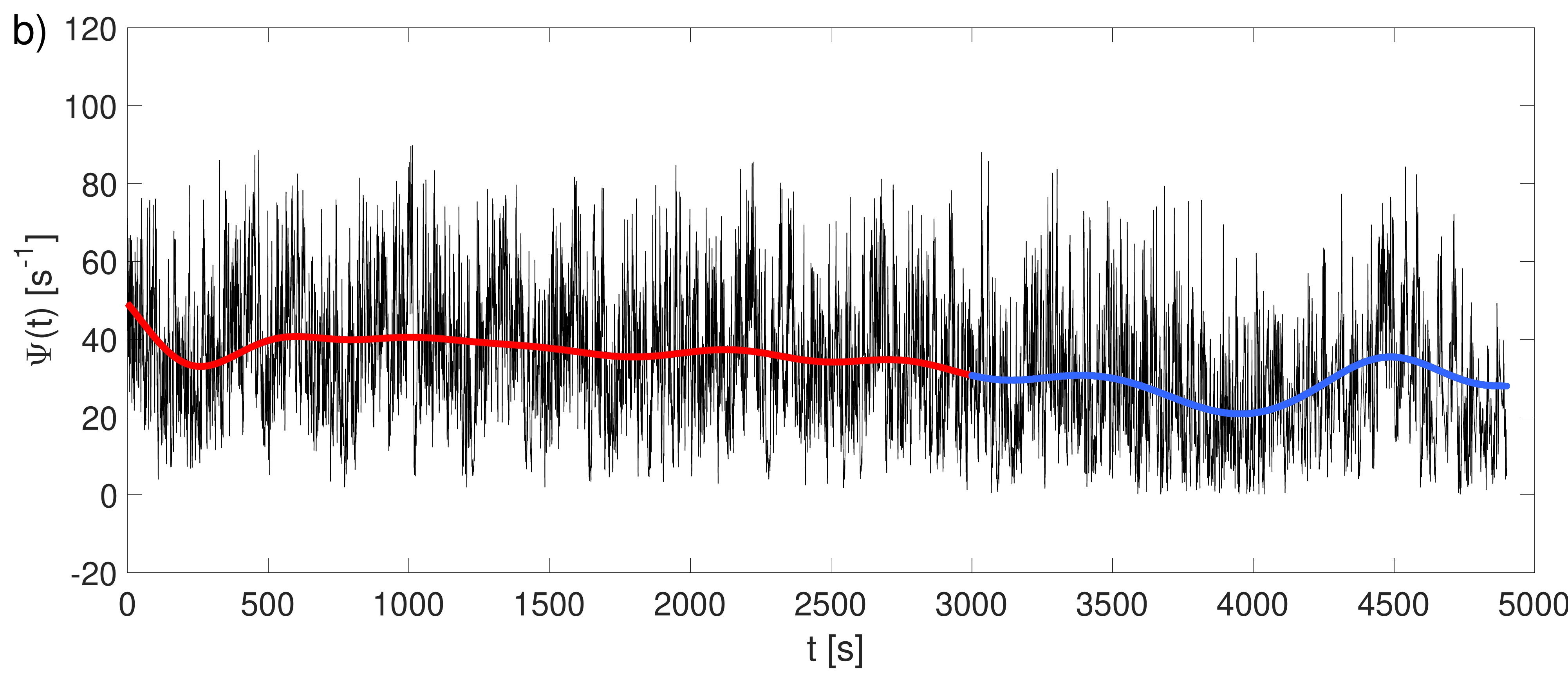} 
\caption{a) Plot of direction $\Phi(t)$ and b) magnitude $\Psi(t)$ of the wall shear-stress vector $\boldsymbol{\tau}(t) / \mu$ over a period of 1.5 hours. Overlaid in color is the profile of the low-pass time-filtered signal of the direction $\widetilde{\Phi}(t)$ and the magnitude $\widetilde{\Psi}(t)$. Two different characteristic phases are coded in color (phase A in red, phase B in blue).}
\label{fig:Zeitverlauf}
\end{figure}

\begin{figure}
\makebox[8cm]{\includegraphics[width=8cm]{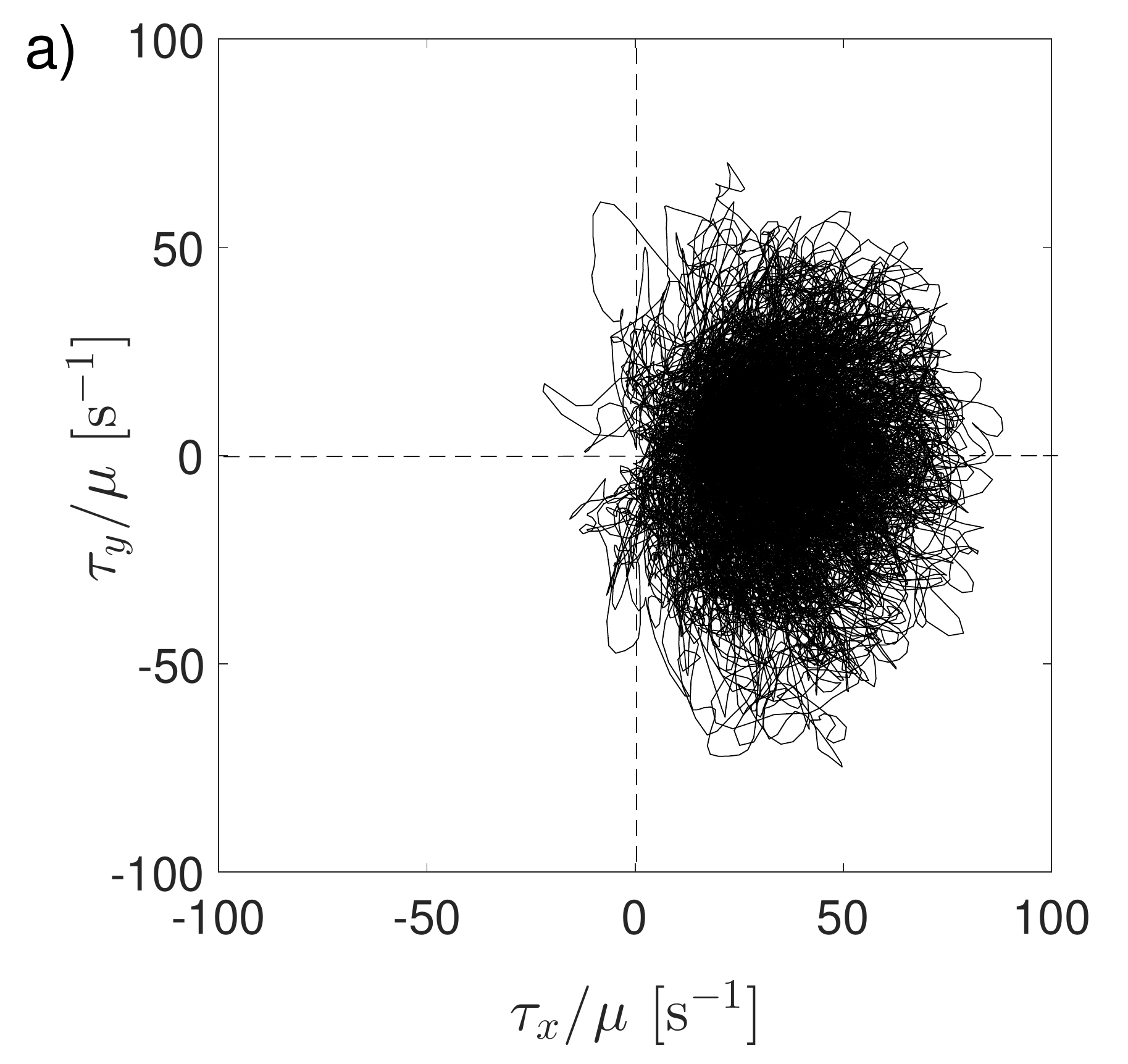}}
\makebox[8cm]{\includegraphics[width=8cm]{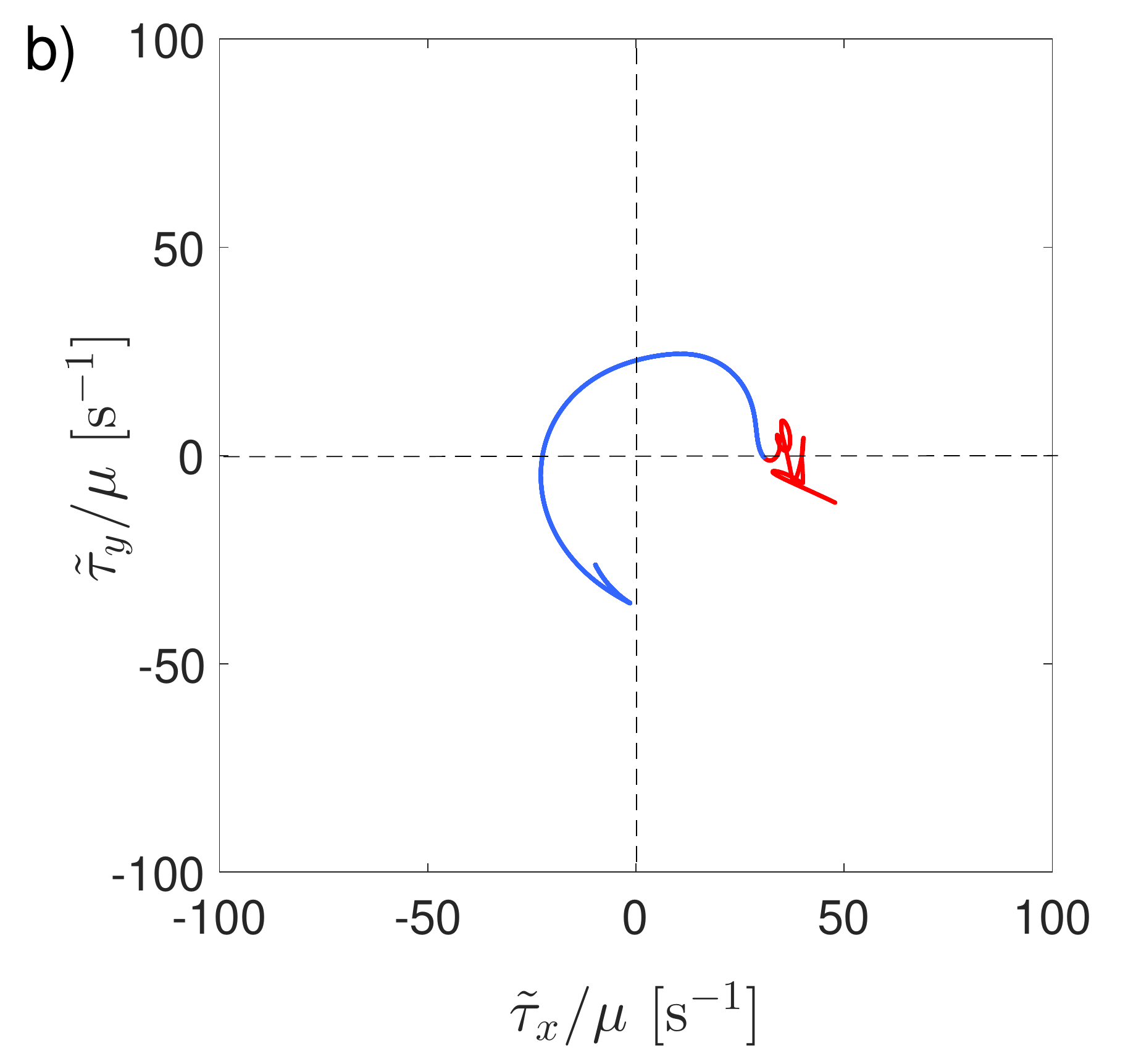}} 
\caption{Temporal trace of the wall shear-stress vector $\boldsymbol{\tau}(t)/\mu$ in the x-y plane, comparable to the trace $\textbf{Q}(t)$ of the sensor head. a) original signal in phase~A; b) low-pass filtered signal in phase~A (colored in red) and in the successive phase~B (colored in blue), compare Fig. 7.}
\label{fig:Lissayou}
\end{figure}

Fig.~\ref{fig:Zeitverlauf} shows time traces of the direction ${\Phi}$, $\widetilde{\Phi}_{LSC}$ and the magnitude $\Psi$, $\widetilde{\Psi}_{LSC}$ over the entirely recorded time span of 1.5~hours.  It is seen that both, direction and magnitude vary strongly. We recognize a fast variation of the orientation of the angle over a range of approximately $\ge~\pm~25{\degree}$, similar as already observed in Shi et al. \cite{Shi12}. On average, the LSC is  almost perfectly aligned with the x-axis for phase A (t=0...3,000~s). A very slow drift of the angle beginning at $t=3000~\textrm{s}$ in phase B (t=3,000...5,000~s) in counter-clockwise direction is superimposed on the fast oscillations. This drift indicates a very slow precession of the LSC. Such a slow precession mode can be present in the BOI, since the mean orientation of the roll is not locked in one particular direction. While the angular fluctuations increase in phase B, the mean magnitude slows down and exhibits a modulation with a long wavelength. Thus, the variance in the pattern of fluctuations increases in phase B relative to the mean and the effect of perturbations accumulates. The mean of the magnitude in phase A amounts to $\overline{\Psi}_A = 40~s^{-1}$ but it significantly decreases in phase B to a value of about $\overline{\Psi}_B= 28~s^{-1}$ (see in Fig.~\ref{fig:Zeitverlauf}). The slow-down of the mean wind lasted about 1,500~s (550 units of $T_f$), meanwhile, the angle $\widetilde{\Phi}_{LSC}$ changed by 1.36 pi. As the kinetic Energy $\overline{E}_{kin}$
is proportional to the square of the magnitude of the wall shear stress  $E_{kin}\propto \Psi^2$, the average kinetic energy of the mean wind in phase B is reduced to about 50~\% of the energy in phase A. In spatially extended and/or high dimensional systems, transitions are often a consequence of such a critical slowdown. Therefore, the lower level of the kinetic energy of the mean wind may have triggered this slow precession. In order to illustrate the complex behaviour of the flow in the x-y plane, the data in Fig.~\ref{fig:Zeitverlauf} are plotted again as a x-y trace plot in Fig.~\ref{fig:Lissayou}.
In phase~A, the direction of the LSC (the red part of the time-filtered signal in Fig.~\ref{fig:Lissayou}b) was almost constant towards north (positive x-axis) with relatively small fluctuations. Later on, in phase~B, the direction of the mean flow followed a nearly circular trace in counter-clockwise direction, which proves that the LSC started to rotate at the time, when the magnitude of the main wind slowed down. Our sensor enables us to reveal this slow precession and to distinguish it from a phase of constant orientation of the wind. The time scale of this precession is rather long compared to the characteristic eddy turnover time of the LSC in the experiment (about $T_e=50~\textrm{s}$). Another  interesting feature in phase~A is the observation of a negative streamwise wall shear stress $\tau_x$ as seen in the traces in Fig.~\ref{fig:Lissayou}, when the line crosses the 2nd or 3rd quadrant in the left sub-figure. Our data indicates that eventually the LSC becomes weaker such that the superposition of strong fluctuations can lead to a local backflow with a negative streamwise wall shear-stress $\tau_x$. Such events are observed in turbulent RB convection for the first time and they are quite similar to those observed in turbulent boundary layer flows \cite{Bruecker2015}.

\begin{figure}
\makebox[8cm]{\includegraphics[width=8cm]{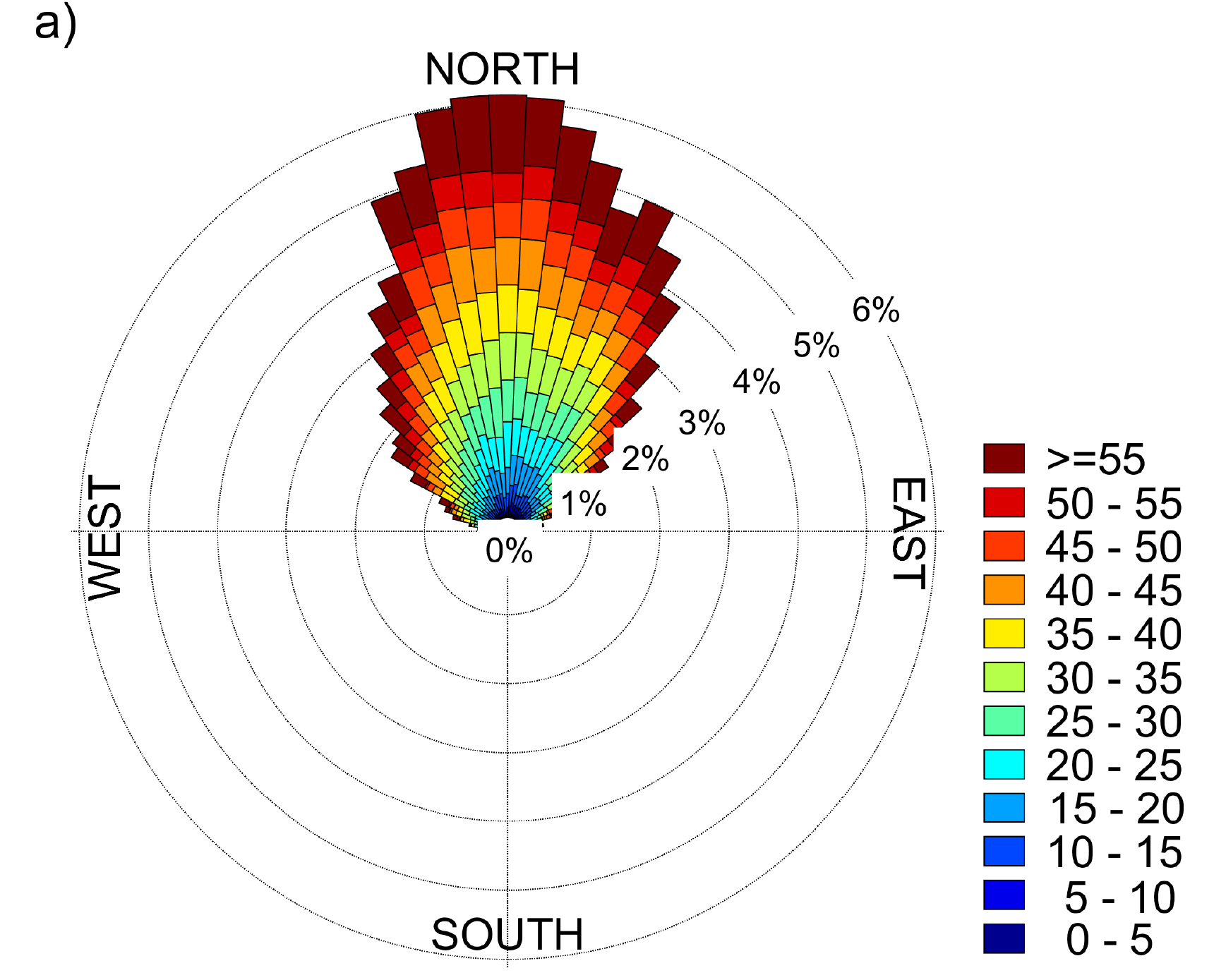}}
\makebox[8cm]{\includegraphics[width=8cm]{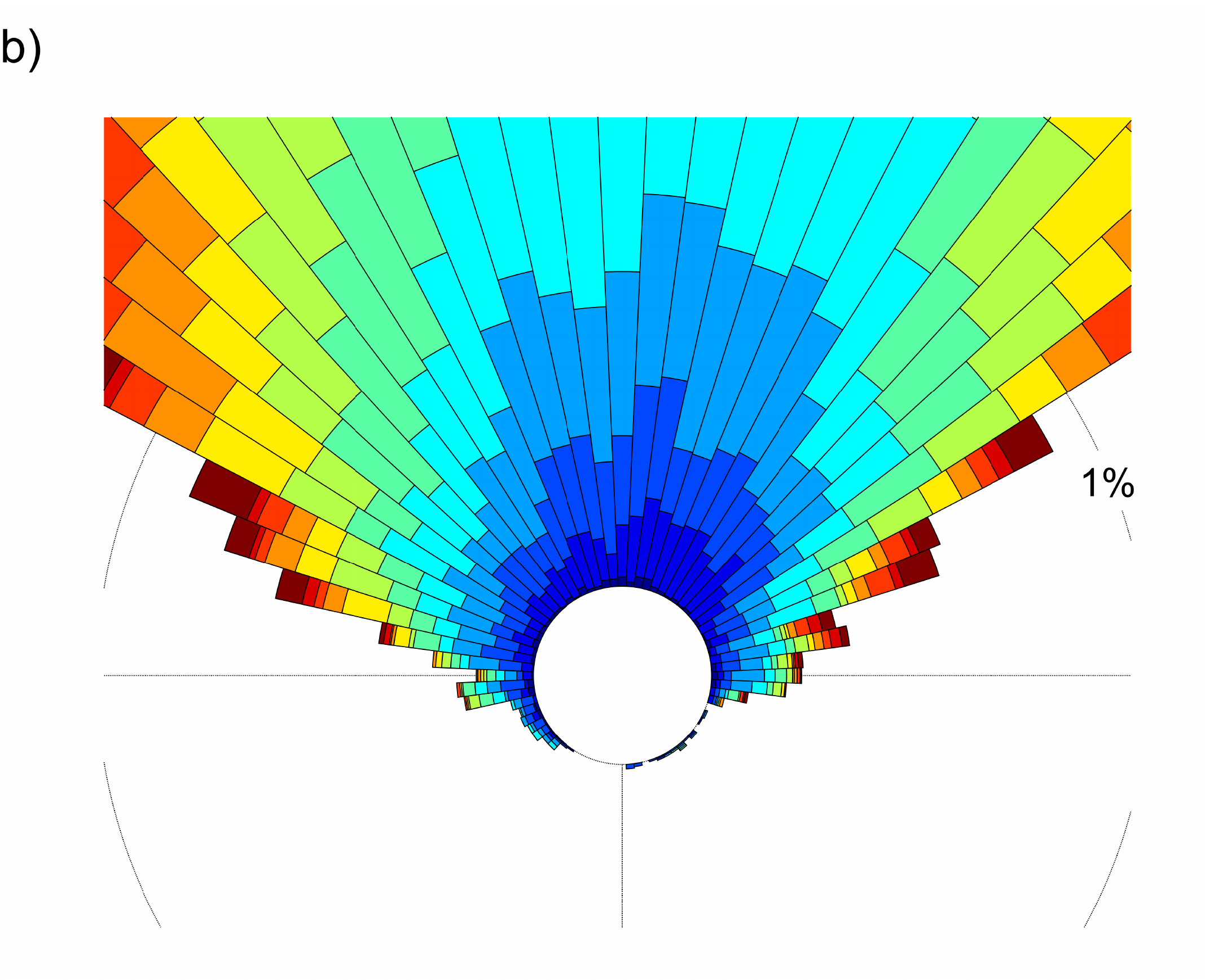}} 
\caption{Polar plot of probability distribution of the wall shear-stress direction $\Phi$ in phase~A (mean wind flow is in direction north). The right sub-figure shows a zoom-in image of the data given left. Angular steps are in 5$\degree$. The color indicates the magnitude $\Psi$ in $s^{-1}$ for the ranges given in the legend bar.}
\label{fig:Windrosen}
\end{figure}

For the further statistical analysis of the wall shear stress in a wind-driven convective boundary layer, we exclusively focus on phase A. In this period, the mean direction of the wind is aligned with the x-axis.  Fig.~\ref{fig:Windrosen} shows the angular probability density function of the yaw angle of the wall shear-stress as a wind rose plate with a mean direction towards north. The angle ${\Phi}$ of the rays relative to north represent the yaw angle, while the length indicates the probability over all samples recorded in phase A. The magnitude $\Psi$ is overlaid in color. The graph is similar to that used by Bruecker displaying the measurements of the statistics of the wall shear stress in TBL flows \cite{Bruecker2015}.  The distribution shows a type of cone, in which mean angles between $\pm~25{\degree}$ around the x-axis predominate. However, there are also, even rarely, events of $\tau$, in which the yaw angle exceeds $\pm~90{\degree}$. The maximum amounts to about  $\pm~110{\degree}$. Thus, these rare events can be associated with events of large spanwise $\tau_y$, first argued in \cite{Bruecker2015} in a zero pressure gradient TBL flow. The probability and the yaw angle of the rare events in thermal convection are quite similar to those reported therein. Therefore, it can be concluded that the source and the character of these events is similar to the transport process in a TBL, where the convection of larger coherent vortex structures within the boundary layer leads to the same effect.

\begin{figure}
\makebox[8cm]{\includegraphics[width=9cm]{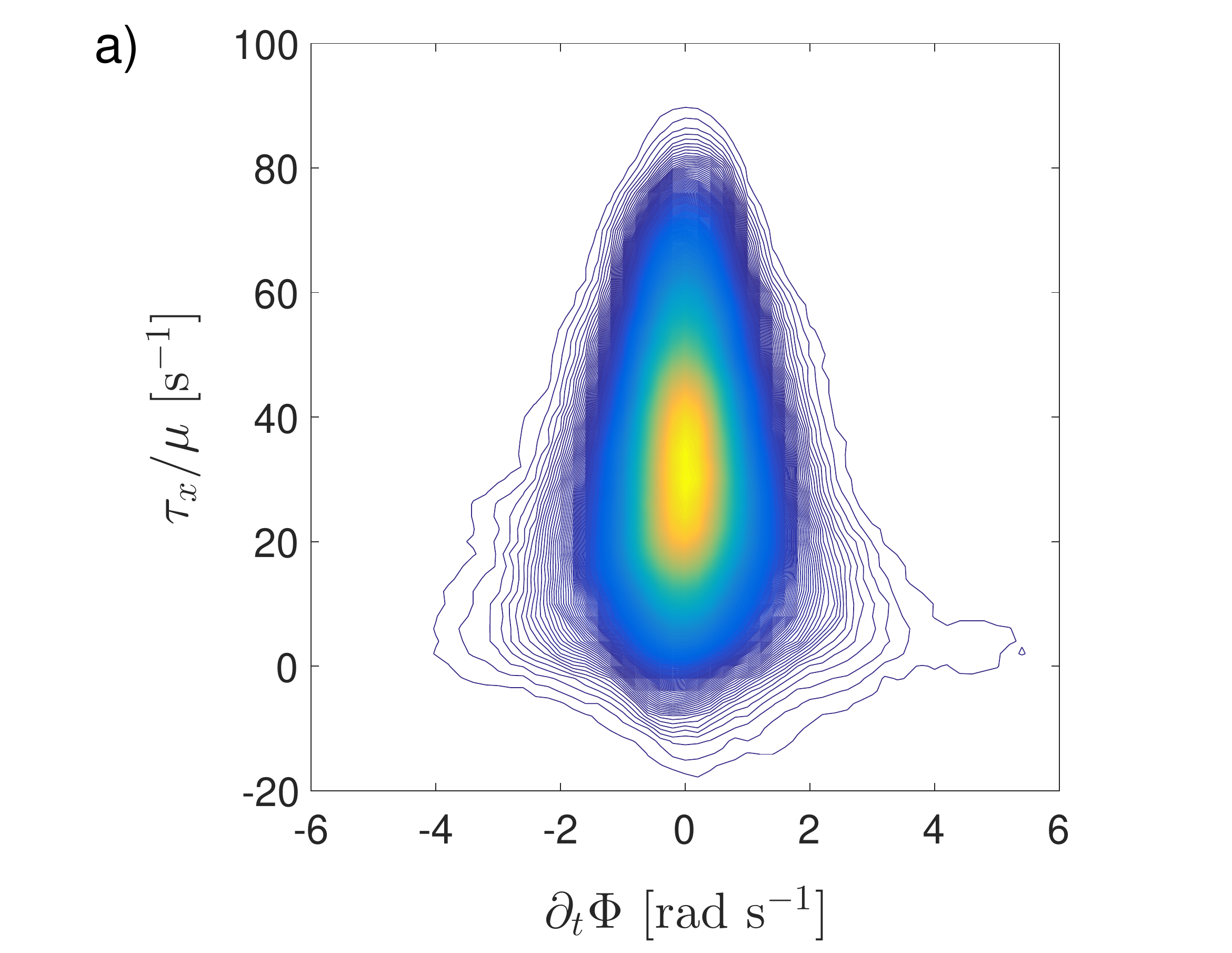}}
\makebox[8cm]{\includegraphics[width=9cm]{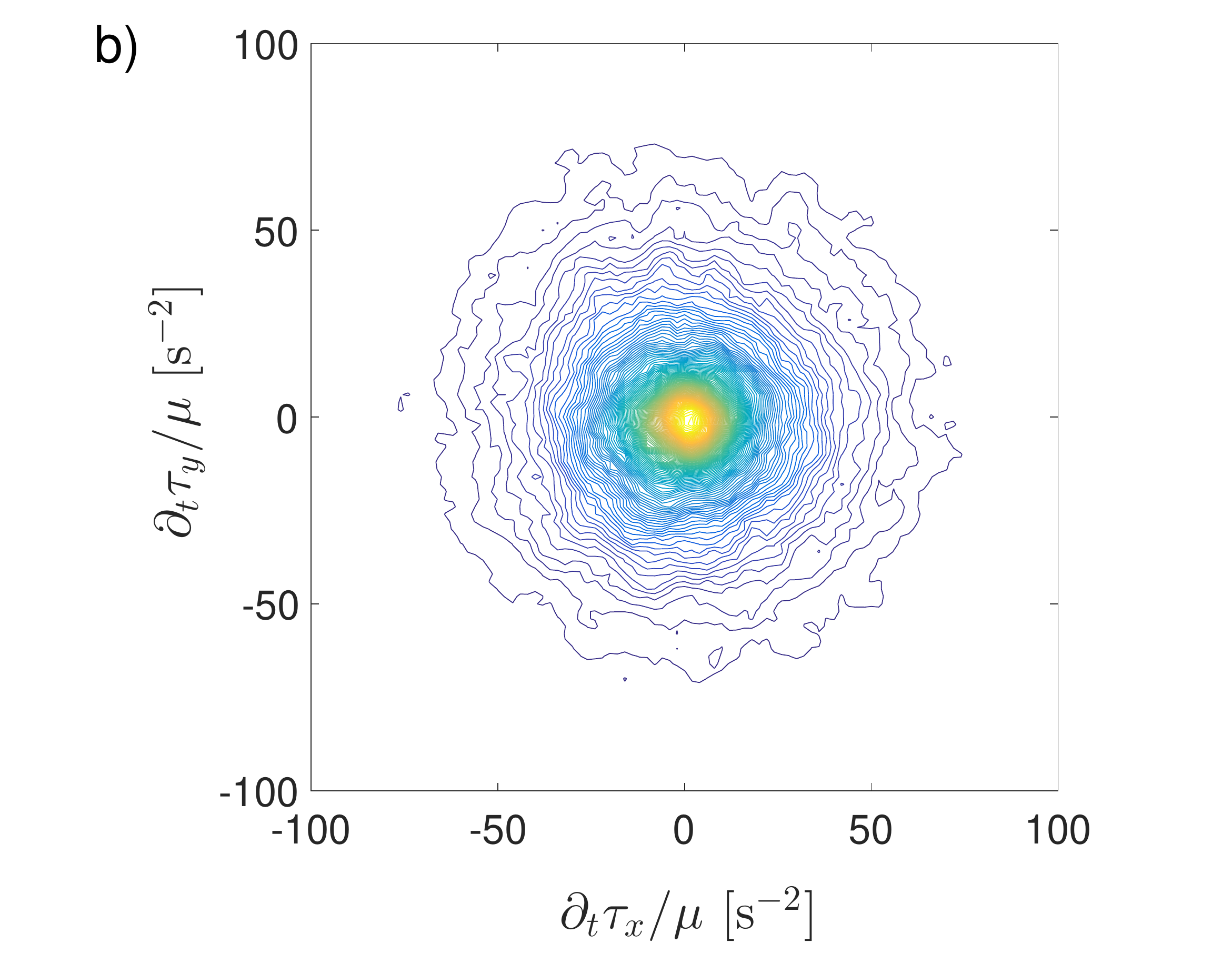}} 
\caption{a) Joint probability density distribution of the streamwise component $\tau_x$ with angular velocity of the wall shear-stress vector; b) joint probability density distribution of temporal gradients of $\tau_x$ and $\tau_y$.}
\label{fig:pdf_acceleration}
\end{figure}

The observation also indicates a rapid temporal variation of the local direction of the fluctuating wall shear-stress, expressed by the angular velocity of the vector $\boldsymbol{\tau}(t)$. We further analyse, how the streamwise wall shear-stress $\tau_x$ correlates with these temporal changes of $\Phi(t)$ by plotting a joint probability density distribution (see Fig.~\ref{fig:pdf_acceleration}a). The colored contour distribution clearly shows the preferred orientation of large positive amplitudes of $\tau_x$ in the direction of the mean-flow with only slow temporal variation in $\Phi$. On the other hand, fluctuations that induce negative streamwise $\tau_x$  relative to the mean wind are correlated with higher angular velocities of the wall shear stress vector. Therefore, the contours display a tear-drop like shape with the tip aligned with the mean wind. A possible explanation for this different behaviour is the passage of two essential elements of the boundary layer: they are documented in previous research as a) the plume-detachment event and b) the post-plume-detachment phase. If a plume dissolves from the boundary layer, the structure of the latter one reorganizes. Typical sequences of such an event cover a certain time lag of about $0.45T_f$. Prior to the plume detachment, the flow is more irregular and hot fluid starts to rise up, probably causing more irregular fluctuations and a twist in $\boldsymbol{\tau}$, reducing  $\tau_x$ and increasing $\tau_y$. If there is any wall-normal vorticity, then, it is reinforced by vortex stretching. This might explain the increase in $\partial_t \Phi$ as a speed-up of the twist observed in this process, similar as in a tornado. This process is represented by the bottom of the tear-drop like shape. Once the plume has formed and rises up, it causes strong upward outflow that is connected to the plume detachment. This is followed with a strong inflow in the back of the plume due to conservation of mass. This is the phase when the fluctuations drive the flow towards the tip of the tear-drop shape. Some more information is given by the temporal rate of change of the spanwise and streamwise component of the wall shear-stress, again displayed as a joint probability density distribution in Fig.~\ref{fig:pdf_acceleration}b. The rate of change of the wall shear-stress can be interpreted as the acceleration of the flow in the plane above the wall. This distribution has a rather isotropic shape, which indicates that there is no preferential direction in flow acceleration. Thus, local pressure gradients are expected to be distributed uniformly in the horizontal plane.

\begin{figure}
	\makebox[8cm]{\includegraphics[width=8cm]{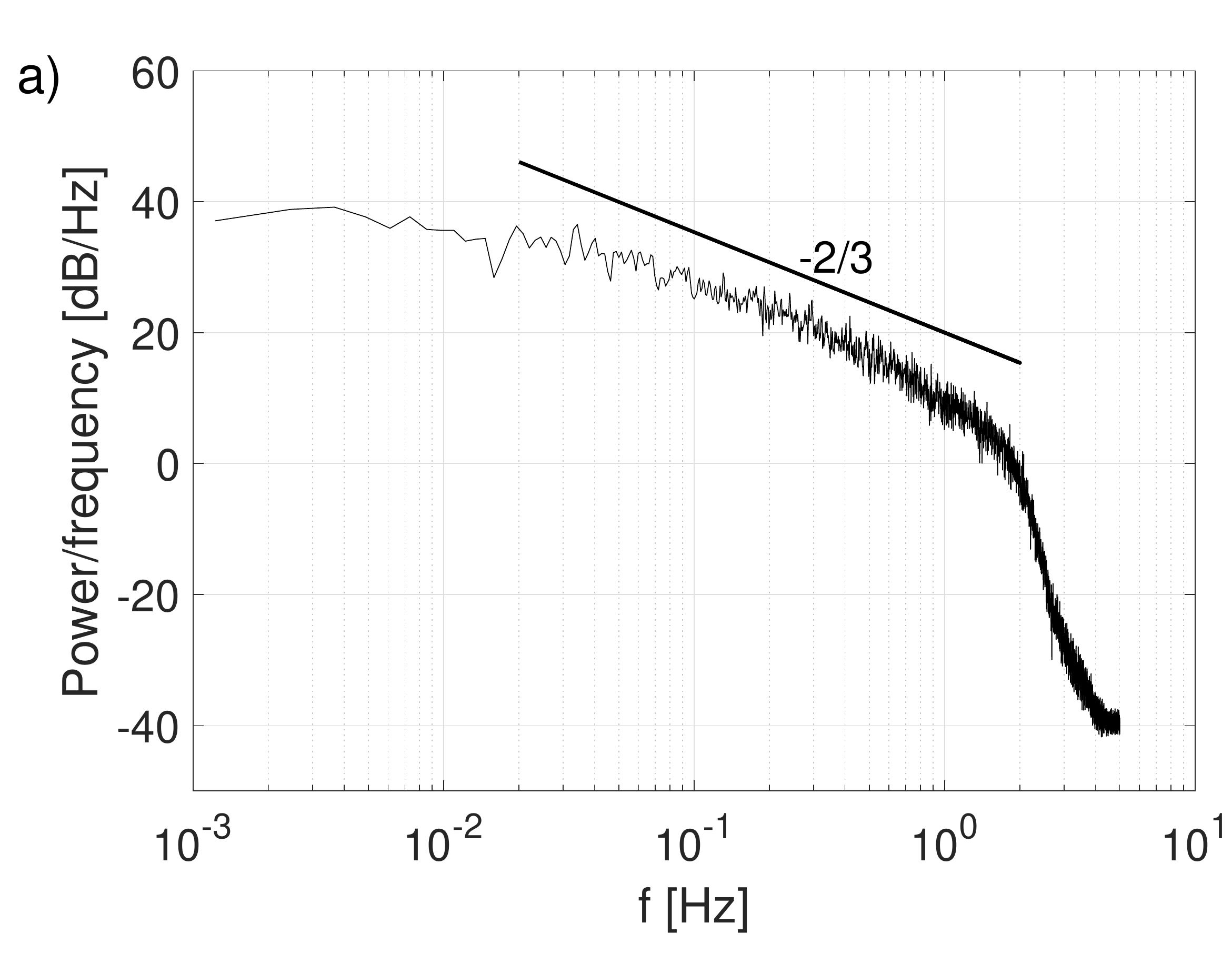}}
	\makebox[8cm]{\includegraphics[width=8cm]{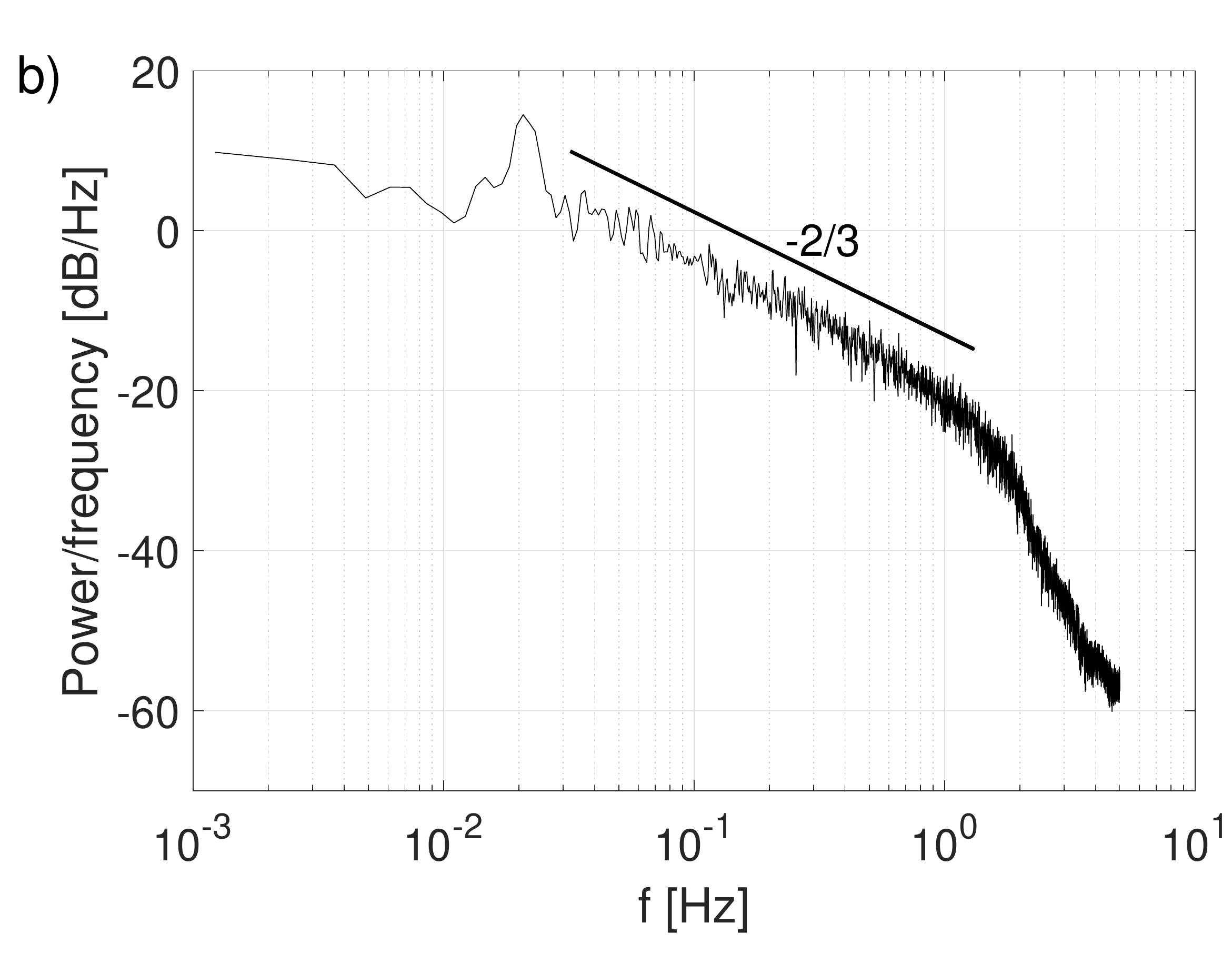}}
	\caption{Power spectra of the magnitude (a) and the angle of the wall shear-stress fluctuations (b). The spectra are calculated using the Welch power estimate \cite{MATLAB:2016}. The roll-off above 2~Hz is caused by the cut-off frequency of the low-pass filter.}
	\label{fig:spektren}
\end{figure}

We further computed the power spectra of the magnitude $\Psi$ and the angle $\Phi$ and the results are given in Fig.~\ref{fig:spektren}. The spectrum in Fig.~\ref{fig:spektren}b clearly shows a prominent peak at about 0.02~Hz. The corresponding time well agrees with the characteristic turnover time $T_e$ of the LSC that has been measured in the BOI in previous experiments \cite{Li12}.  It represents the angular oscillation of the LSC, which is not visible in the spectrum of the magnitude in Fig.~\ref{fig:spektren}a. Both spectra follow an approximately linear decay in the log-log plot for the range of 0.03--1~Hz with a slope of about -2/3. Since the sampling frequency is as high as 10~Hz, the spectrum ends up at 5~Hz. As mentioned above, we filtered our signal by a low-pass filter with a cut-off frequency of 2~Hz. The steep roll-off of the spectrum beyond this limit follows the natural decrease of the forth-order butter-worth filter (-80 dB/decade) and is not associated with any physical process in the flow.

\begin{figure}
\makebox[8cm]{\includegraphics[width=8cm]{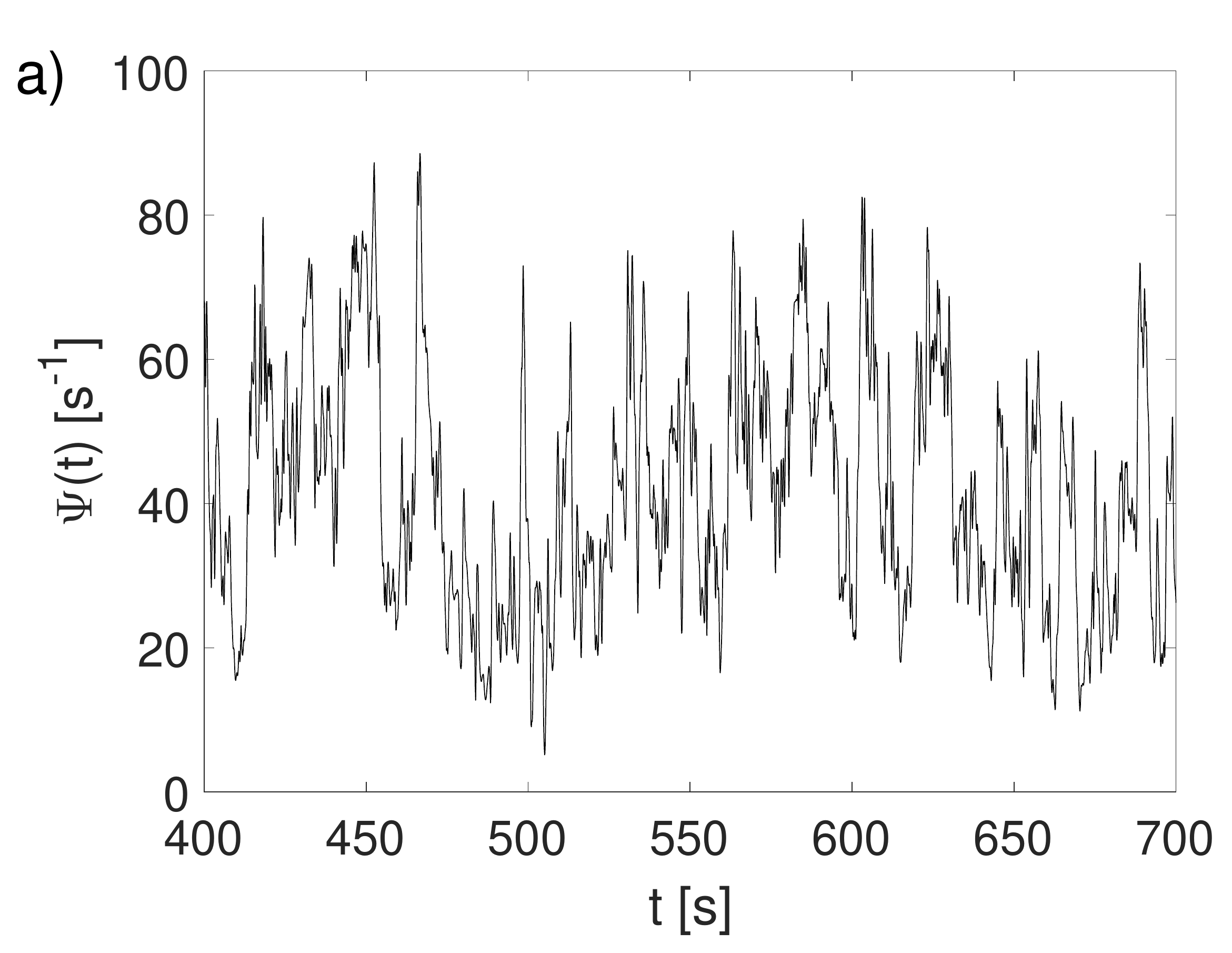}}
\makebox[8cm]{\includegraphics[width=8cm]{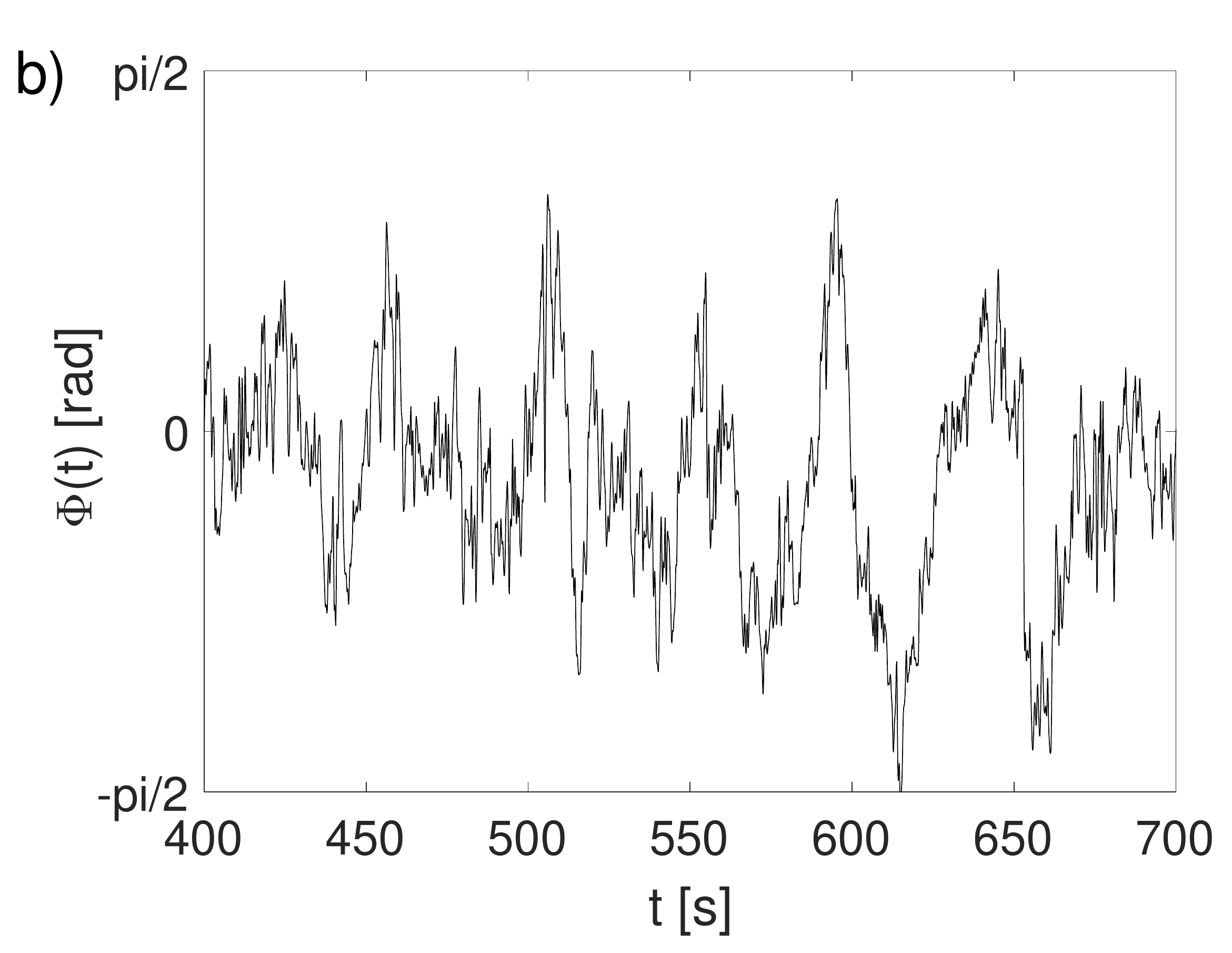}}
\makebox[8cm]{\includegraphics[width=8cm]{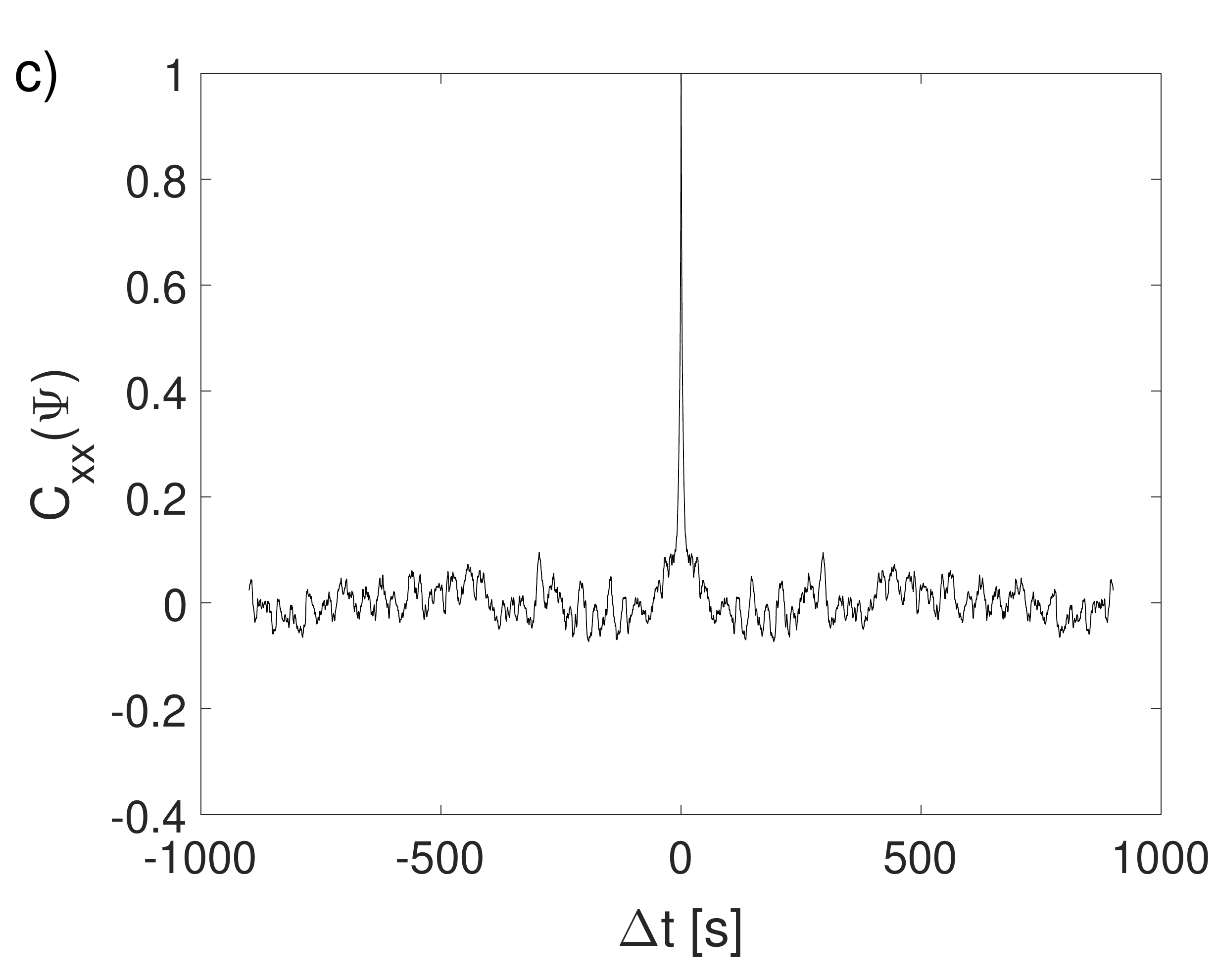}}
\makebox[8cm]{\includegraphics[width=8cm]{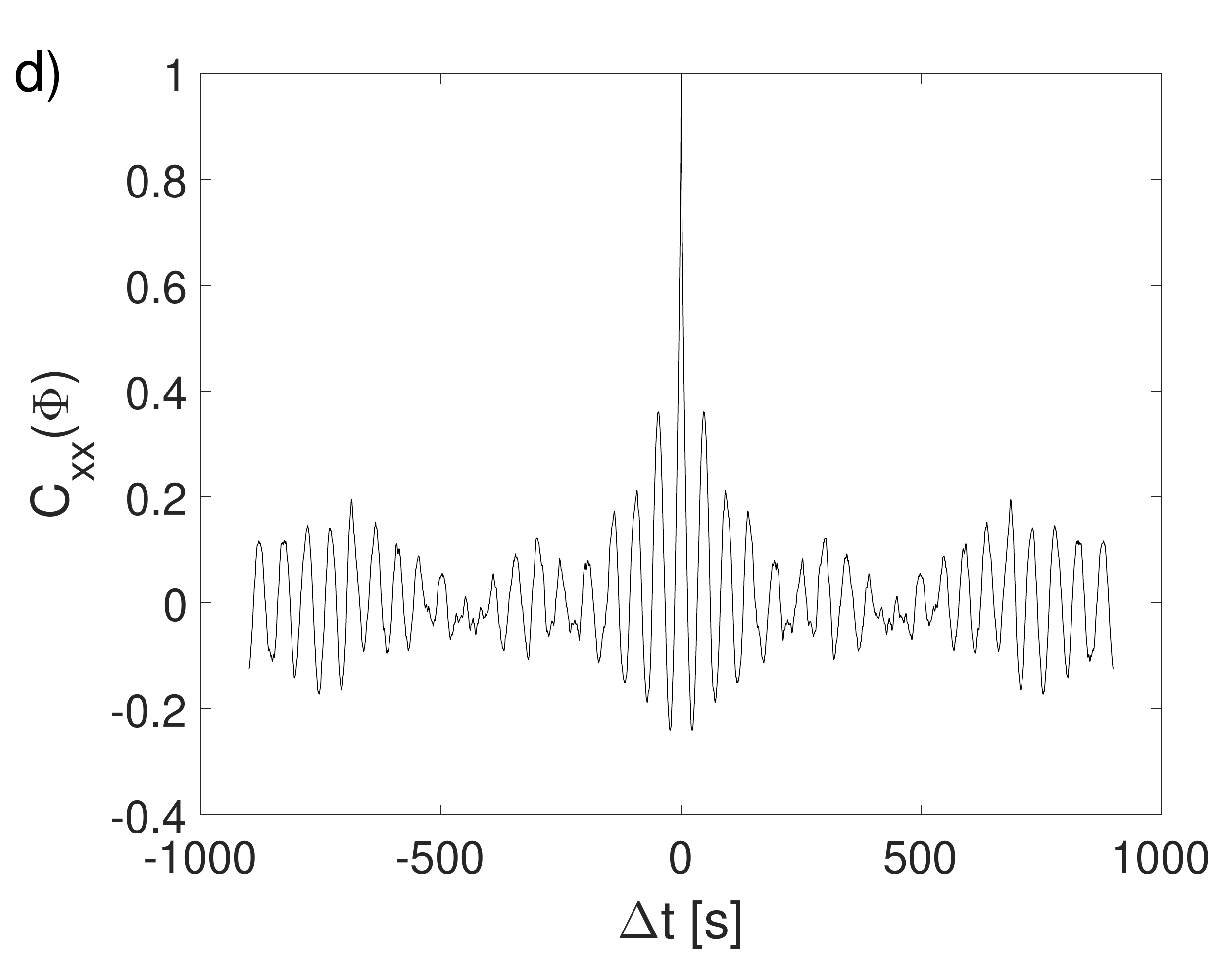}}
\caption{Time series of magnitude $\Psi(t)$ and direction $\Phi(t)$ of the wall shear-stress vector $\boldsymbol{\tau} /\mu$ for phase~A, $400~\textrm{s}<t<700~\textrm{s}$ (a and b); autocorrelation function of the magnitude $C_{xx}(\Psi)$ and the direction $C_{xx}(\Phi)$ for phase~A (c and d).}
\label{fig:signals_and_autocorrelation}
\end{figure}

In order to analyse the angular oscillation of the LSC in more detail, one each short window of the time traces of $\Psi(t)$ and $\Phi(t)$ are plotted in detail in Fig.~\ref{fig:signals_and_autocorrelation}. While the oscillations of the magnitude seem to be rather irregular (see Fig.~\ref{fig:signals_and_autocorrelation}a), the plot of $\Phi(t)$ reveals a low frequency oscillation around the mean with a frequency of about 0.02~Hz (see Fig.~\ref{fig:signals_and_autocorrelation}b). This timescale corresponds to the characteristic turnover time $T_e$ of the LSC. It is more pronounced in the plot of the autocorrelation function $C_{xx}(\Phi)$ and $C_{xx}(\Psi)$ (see Figs.~\ref{fig:signals_and_autocorrelation}c,d). The autocorrelation $C_{xx}(\Phi)$ shows that the system of the LSC is in a quasi-bistable state. We conclude this from the regular angular fluctuations (Fig.~\ref{fig:signals_and_autocorrelation}b) and the strong periodic correlation peaks. The dynamical system has obviously two attracting states overlaid with a certain fraction of noise/turbulence. This is quite similar to other bistable systems as, \textit{e.g.}, reported in \cite{Kumar13}. Consequently, the RB convection system also exhibits two time scales: the periodic dynamics representing the low frequency angular oscillations of the mean wind and turbulent small-scale fluctuations. Another observation is the long-term modulation of the angular oscillations as the peak amplitude of the correlation slowly decreases for increasing time lag to about zero at a time lag of $\triangle t=450~\textrm{s}$ ($9T_e$). For even higher $\triangle t$, it increases again to a correlation value $C_{xx}(\Phi)=0.2$. The second maximum appears at $\triangle t=700~\textrm{s}$ ($14T_e$). This decrease in autocorrelation towards a minimum can be explained applying Extreme Value Theory (EVT). According to the ideas of EVT the decrease of the correlation towards a minimum is associated with a critical slow down, in which the system becomes increasingly weak, while recovering from small perturbations \cite{Kumar13}.
\begin{figure}
\makebox[16cm]{\includegraphics[width=16cm]{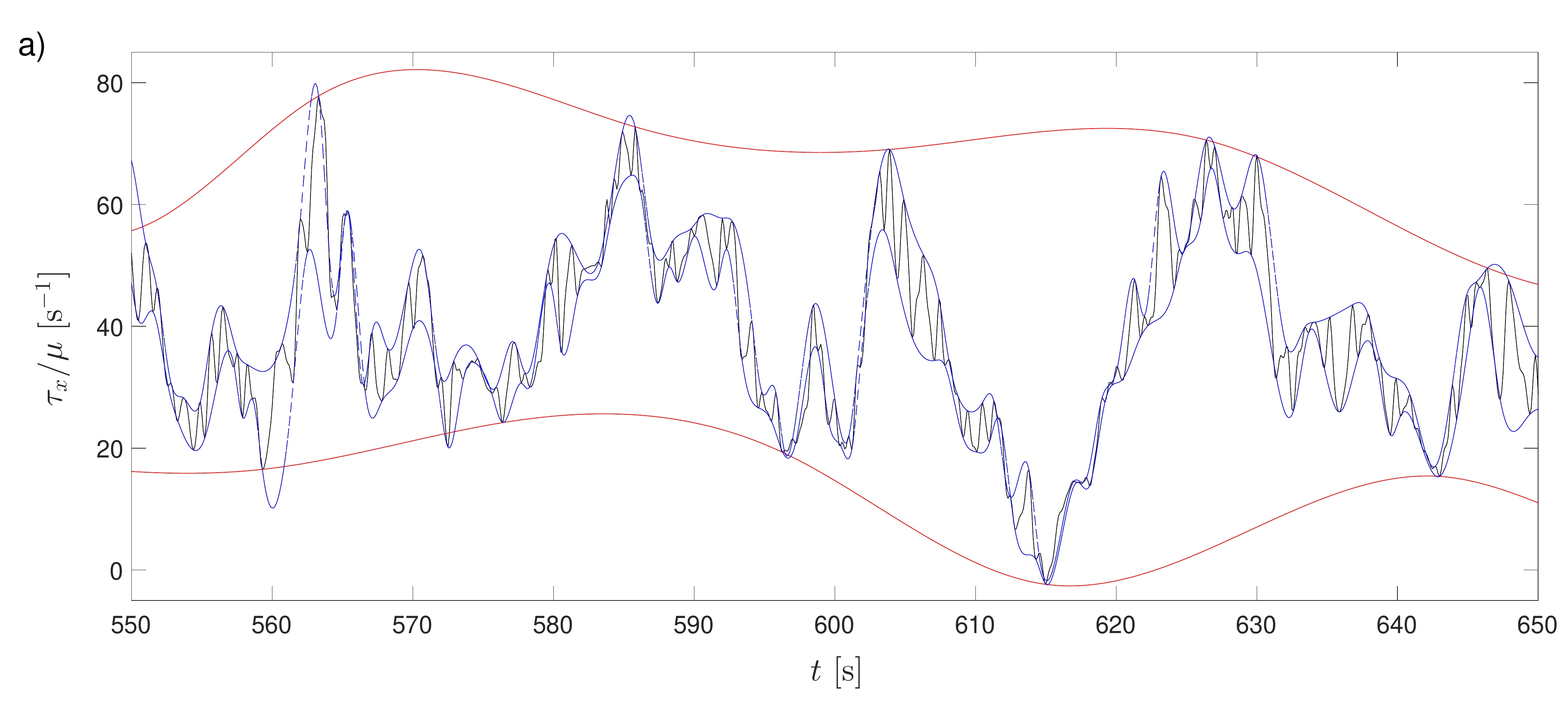}} 
\makebox[8cm]{\includegraphics[width=7.7cm]{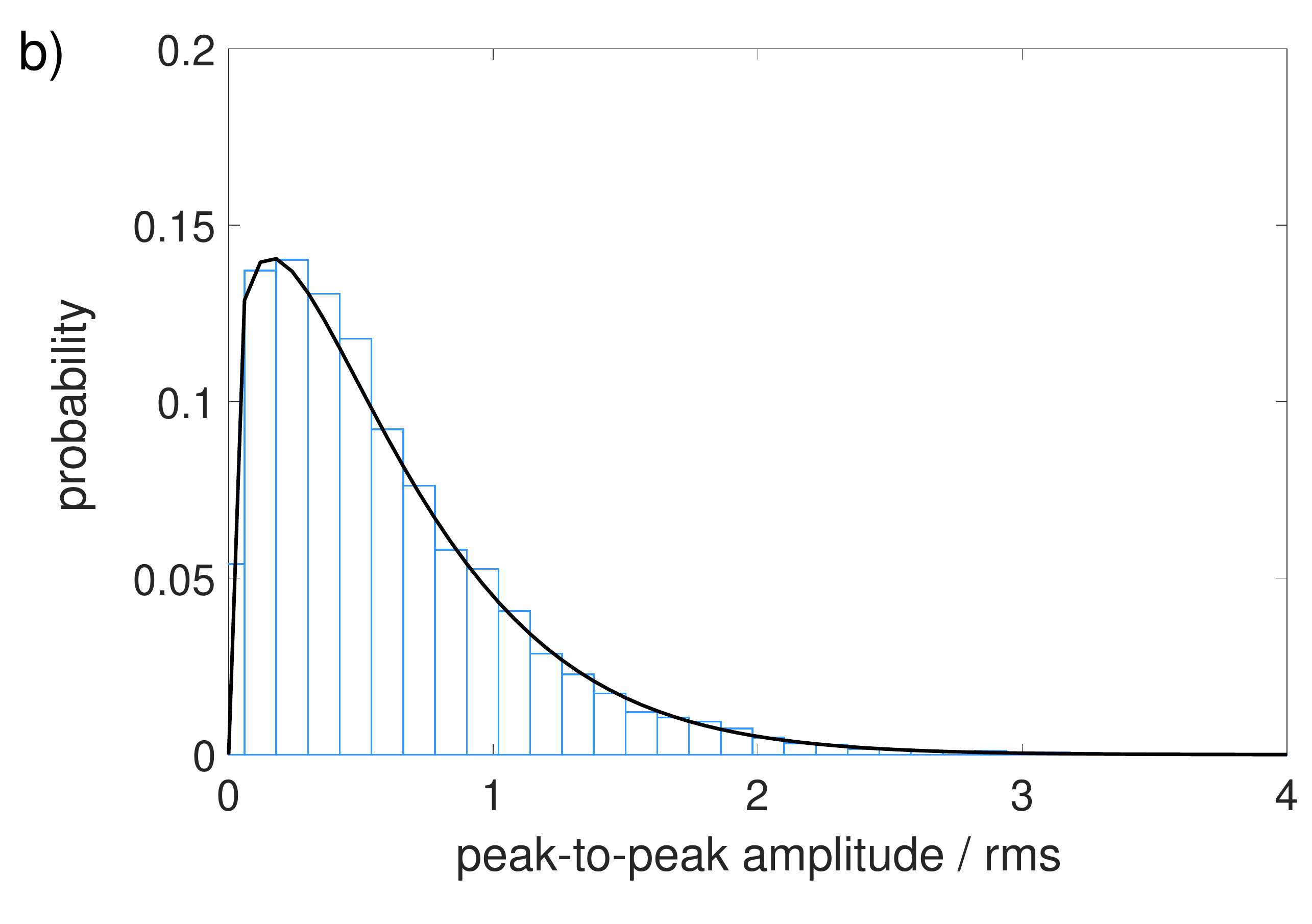}}
\makebox[8cm]{\includegraphics[width=7.7cm]{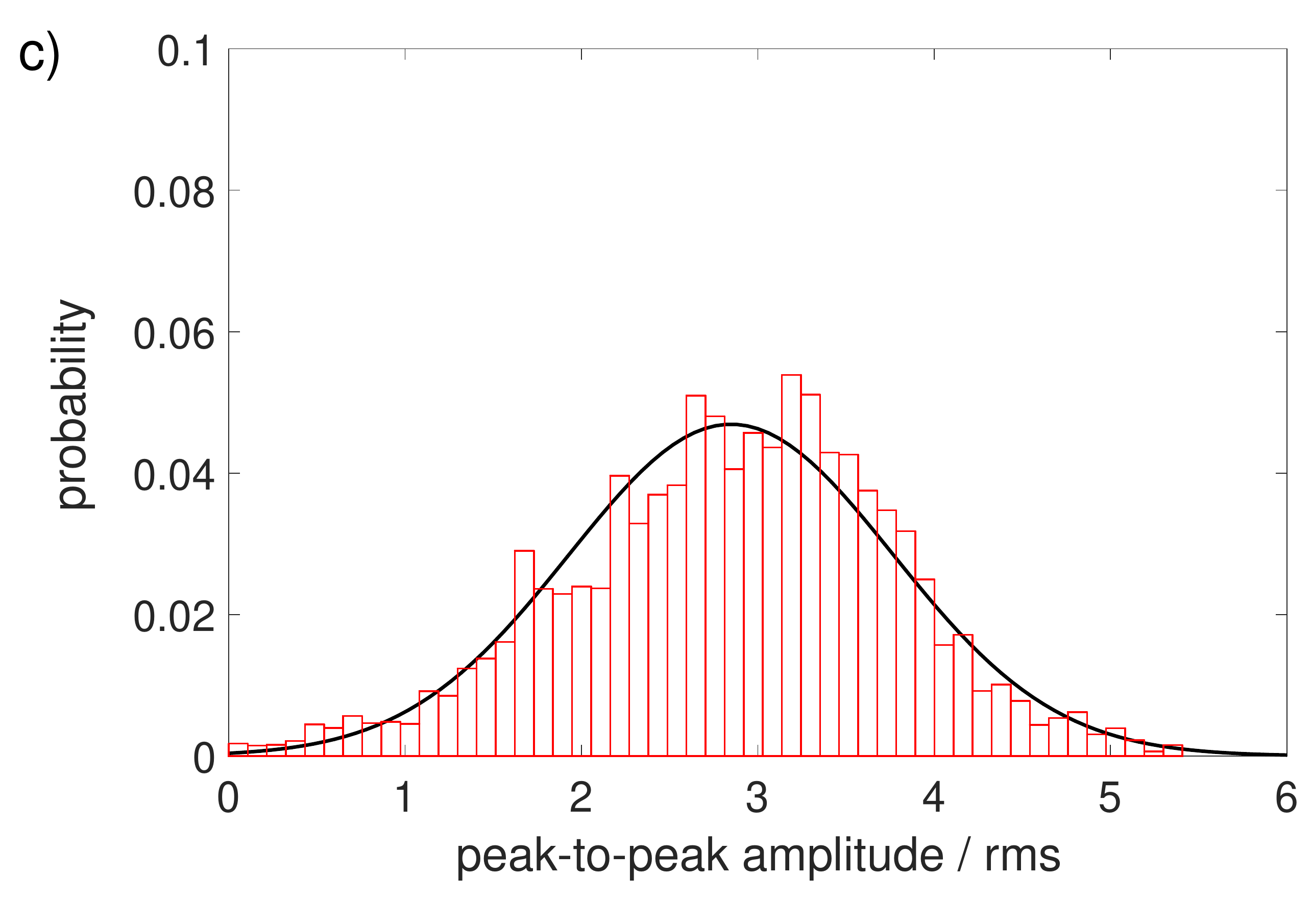}}
\caption{Peak-to-peak amplitude of the fluctuations of $\tau_x$ at different timescales. a) Temporal signal $\tau_x$ (black line) with envelopes of the short-term and long-term fluctuations (thin blue and red lines, respectively). b) Probability density function (PDF) of the short-term fluctuations (the solid line is a Weibull fit with a scale parameter $\lambda=0.637$ and a shape parameter $k=1.223$), c) PDF of the long-term fluctuations (the solid line is a Gaussian fit with a mean value of $\bar{\tau}_x=2.85$ and a standard deviation of $\sigma(\tau_x)=0.92$).}
\label{fig:envelope}
\end{figure}
It is quite difficult to recover extreme events using classical conditional averaging methods or a fixed threshold definition due to this particular modulation of the magnitude and the orientation of the mean flow. Here, we discriminate the oscillation of the mean flow into different time-scales. We distinguish between the periodic transitive dynamics represented in the low-frequency angular oscillations of the mean wind and the small-scale turbulent fluctuations. To this end, we apply envelope functions with different time windows on $\tau_x$ to determine the amplitudes of these fluctuations on the different timescales. The envelope is calculated from the Matlab toolbox and uses a sliding time-window that connects within the window the local peaks (upper envelope for local maxima and lower envelope for local minimum peaks) with a smoothed spline \cite{MATLAB:2016}. For the low frequency dynamics, we chose a window of 15~s, while using a shorter time window of 0.5~s for the small-scale turbulent structures. One typical example of such an enveloping curve is plotted along with the original signal in Fig.~\ref{fig:envelope}a. In order to analyse the amplitude of the fluctuations, we compute the absolute difference between the upper and the lower envelopes and determine the probability density function (PDF) for both time windows (see Fig.~\ref{fig:envelope}b and c). The PDF of the short-term fluctuations using a time window of 0.5~s is shown in Fig.~\ref{fig:envelope}b, that for the long-term fluctuations is shown in Fig.~\ref{fig:envelope}c. While the short-term fluctuations of the streamwise wall shear-stress follow a Weibull distribution according to:
\begin{equation}
f(x;\lambda, k)=\frac{k}{\lambda}(\frac{x}{k})^{k-1} e^{-(x/\lambda)^k}
\label{eq:Weibull}
\end{equation}
(with the scale parameter $\lambda$=0.637 and the shape parameter $k$=1.223), the long-term fluctuations are clearly Gaussian distributed. In conclusion, extreme events are more likely, if large excursions occur simultaneous for both statistical distributions. Our long-term recording of totally 54,000 samples covers more than 100 LSC turnover times $T_e$, and ensures sufficient statistical evidence even for the long time-scales.

\section{\label{sec:Conclusion}Conclusion}

We have investigated the long-term behavior of the wall shear-stress fluctuations in turbulent Rayleigh-B\'enard convection in air. Using a novel sensor concept based on a nature-inspired pappus design, we measured the direction and the magnitude of the wall shear stress vector $\boldsymbol{\tau}(t)$ at the bottom plate of a large-scale convection cell, the ``Barrel of Ilmenau''. For a fixed cylindrical geometry of aspect ratio unity and a Rayleigh-number of $Ra=1.58\times10^{10}$, we recorded the fluctuations using an optical method, and we analysed their angular distribution at different timescales. The wall shear-stress vector $\boldsymbol{\tau}(t)$ measured at the center of the heated bottom plate clearly reflects the strength and the orientation of the large-scale circulation (LSC), being omnipresent in such a system. We demonstrate that it remains in a bistable state for a very long time. Its orientation oscillates regularly with $\pm25\degree$ around the mean, which represents the typical wind direction in the convection cell. The period of this angular oscillations is linked to the characteristic turnover time $T_e$ of the LSC in the cell. The analysis of the autocorrelation of the wall shear-stress direction $\Phi$ yields a considerable large memory of those oscillations up to temporal scales of several decades of turnover cycles. The peaks in the correlation plot show also a low frequency modulation of these oscillations, whose timescale agrees with a modulation of the magnitude of the mean wall shear stress, respectively of the mean wind in the cell. After this long phase of constant mean wind direction a transition sets in that causes the slow precession of the mean wind orientation. The periodic oscillations of the LSC persist over this transition while slowly precessing. It is concluded that this transition is triggered by a critical decay in mean wind kinetic energy, which happens, when the observed very low-frequency amplitude modulation in the wall shear-stress reaches a local minimum. On the same time, the level of perturbation remains rather unchanged, thus their influence increases within this phase. Within the regular oscillations of the LSC, extreme events such as local backflow events were observed, seen by negative streamwise wall shear-stress. The latter phenomenon has also been detected in turbulent boundary layer flow along a flat wall \cite{Bruecker2015} recently. This is the first time that such events also could be documented in a temperature gradient driven flow. The similarity in the statistics of the small-scale fluctuations indicates the presence of coherent vortices as characteristic for a turbulent boundary layer, which is generated by the LSC. Our measurements show, that these events are correlated with a rapid twist of the wall shear-stress vector.  As these local flow-reversals are related to singularities in the skin-friction field as shown in \cite{Bruecker2015}, we can follow the argument given in \cite{Bandaru2015} about a possible link of these events to the detachment of thermal plumes.  

We separate the fluctuations of $\tau_x$ in two different timescales (0.5~s and 15~s) using a method that calculates the envelope of the extreme events (peaks of local maxima and minima). The short-term fluctuations show peak-to-peak amplitudes that follow a highly skewed Weibull distribution, which indicates the strong intermittent character of these turbulent fluctuations. On the contrary, for the long-term fluctuations the peak-to-peak amplitudes are represented by a symmetric Gaussian. In both distributions the ends of the tails can reach amplitude values of 3-4 times the rms of the wall shear-stress. A coincidence of large values in both distributions, therefore, may lead to rare excursions such as the reorientation of the LSC or the appearance of local events of a negative streamwise wall shear stress (the latter ones could be validated in our experiment). This behaviour suggests that the fluctuations of the wall shear-stress observed at the bottom wall are a superposition of the long-term dynamics of the bi-stable LSC and the turbulent, highly intermittent near-wall events. Most likely, these are impinging jets from the fully turbulent core of the flow, which transport coherent structures along the wall. All these observations show that the flow close to the heated bottom and the cooled top plates is similar to other dynamical systems with a bi-stable set of states, where fluctuations of considerable strong amplitude trigger the oscillation of the system between both states. The slow precessing of the mean flow direction is a transient phase or a reorientation of the flow in confined thermal convection, while the oscillating flow of the LSC is seemingly persistent during this phase.

The current conclusions on the wall shear-stress measurements built on the fact that the sensor height is one order of magnitude smaller than the boundary layer thickness in the BOI at the given conditions. It is sufficiently small to consider the velocity profile from the wall up to the sensor head as linear in a first approximation. Regarding this, we wish to refer to a very recent work of Daniel \textit{et al.} \cite{Diaz2017}. The authors of this paper studied the correlation between the fluctuations of the wall shear stress and the fluctuations of the velocity in the near-wall region in a zero-pressure-gradient TBL at high Re-numbers. When we apply this analysis to our own set-up, their results predict a correlation coefficient of $C=0.3$. Recalling, that the low-level turbulent natural convection flow near the wall herein cannot be compared one-to-one to the situation of a high Re number TBL along a flat plate. The wall shear stress dynamics in thermal convection is mainly driven by the eruptions of hot plumes or impacts of cold ones. Therefore, we expect in our experiments a higher correlation of the near-wall flow features with the dynamics of the wall shear-stress. Insofar, although the sensor might not be perfect, it is still of sufficient sensitivity and selectivity to provide a first insight into the statistical properties of the different components of the wall shear-stress as well as the dynamics of the local wall heat flux which is, according to Prandtl's mixing length theory \cite{Prandtl1925}, directly linked to the local wall shear-stress.

\begin{acknowledgments}
The position of Professor Christoph Bruecker is co-funded as the BAE SYSTEMS Sir Richard Olver Chair and the Royal Academy of Engineering Chair (grant RCSRF1617/4/11) which is gratefully acknowledged. We wish to acknowledge the support of the European Union under the Grant Agreement number 312778 as well as the support from the German Research Foundation under the grant number PU436/1-2. Moreover, we thank Vladimir Mikulich, Sabine Abawi, and Vigimantas Mitschunas for the technical assistance to run the experiment.
\end{acknowledgments}

\appendix

\bibliography{2018_03_21_dupuits}

\end{document}